\documentclass[structabstract]{aa}  
\usepackage{graphicx}
\usepackage{txfonts}
\usepackage{epsfig,times,natbib,array}
\bibpunct{(}{)}{;}{a}{}{,}

 \newcommand{\chandra}{{\it
    Chandra} }   
\newcommand{\xmm}{{\it XMM-Newton} }
\begin{document}
   \title{Inhomogeneous Metal Distribution in the Intra-Cluster Medium}

%   \subtitle{I. Overviewing the $\kappa$-mechanism}

   \author{L. Lovisari\inst{1}
          \and
          S. Schindler\inst{1}
          \and
          W. Kapferer\inst{1}
% \fnmsep\thanks{Send the offprint requests to L. Lovisari}
          }

   \institute{Institut f{\"u}r Astro- und Teilchenphysik,
  Universit{\"a}t Innsbruck, Technikerstr. 25, A-6020 Innsbruck,
  Austria\\
              \email{Lorenzo.Lovisari@uibk.ac.at}
%         \and
%             University of Alexandria, Department of Geography, ...\\
%            \email{Lorenzo.Lovisari@uibk.ac.at}
%             \thanks{The university of heaven temporarily does not
%                     accept e-mails}
             }

   \date{Received ; }

% \abstract{}{}{}{}{} 
% 5 {} token are mandatory
 
  \abstract
  {The hot gas that fills the space
  between galaxies in clusters is rich in metals. In their large
  potential wells, galaxy clusters accumulate metals over the whole
  cluster history and hence they retain important information on cluster
  formation and evolution.}
  {We use a sample of 5 cool core clusters
  to study the distribution of metals in the ICM. We investigate
  whether the X-ray observations yield good estimates for the metal
  mass and whether the heavy elements abundances are consistent with a
  certain relative fraction of SN Ia to SNCC.}
  {We derive detailed metallicity maps
  of the clusters from $\xmm$ observations and we use them as a measure
  for the metal mass in the ICM. We determine radial profiles for
  several elements and using population synthesis and chemical
  enrichment models, we study the agreement between the measured
  abundances and the theoretical yields.}
  {We show that even in relaxed clusters
  the distribution of metals show a lot of inhomogeneities. Using 
  metal maps usually gives a metal mass 10-30$\%$ higher than the
  metal mass computed using a single extraction region, hence it is
  expected that most previous metal mass determination have
  underestimated metal mass. The abundance ratio of $\alpha$-elements
  to Fe, even in the central parts of clusters, are consistent with an
  enrichment due to the combination of SN Ia and SNCC.}
   {}


   \maketitle
%
%________________________________________________________________

\section{Introduction}
Since the first X-ray observations of the 7 keV iron line feature in
the 1970's by \cite{1976MNRAS.175P..29M} we know that the
intra-cluster medium (ICM) does not only contain primordial elements
but also heavy elements. As heavy elements are only produced in stars
which reside mainly in galaxies the enriched material must have been
ejected into the ICM by the member galaxies. Due to the large
potential wells of galaxy clusters they retain all the enriched
material, so it makes them excellent laboratories for the study of
nucleosynthesis and of the chemical enrichment history of the
universe. Because the gas transfer affects the galaxy and galaxy
cluster evolution, it is important to know when and how the enrichment
takes place.\\ The components in a galaxy cluster interact with each
other in many different ways, thus to study the distribution of the
ejected metals can give us important information on the mechanisms
that transported the enriched gas into the ICM. \\ Several processes
were proposed to explain the observed enrichment in the ICM:
ram-pressure stripping (\citealt{1972ApJ...176....1G}), galactic winds
(\citealt{1978ApJ...223...47D}), galaxy-galaxy interactions
(\citealt{1998MNRAS.294..407G}), AGN outflows
(\citealt{1986ApJ...307...62D}; \citealt{2002ApJ...573L..77H}),
intra-cluster supernovae (\citealt{2002ApJ...580L.121G}) and
others. Simulations show an inhomogeneous distribution of the metals
independent on the enrichment processes
(\citealt{2006A&A...447..827K},
\citealt{2008SSRv..134..363S}). Although AGN outflows as well as
galaxy-galaxy interactions can add metals to the ICM
(\citealt{2005A&A...438...87K}; \citealt{2007MNRAS.374..787H}),
simulations suggest that the metal enrichment of the ICM is primarily
due to galactic winds and ram-pressure stripping. A detailed
comparison between the enrichment due to galactic winds and
ram-pressure stripping revealed that these two processes yield
different metal distributions and a different time dependence of the
enrichment (\citealt{2007A&A...466..813K}).  In massive clusters
ram-pressure stripping provides a much more centrally concentrated
distribution than galactic winds, because galactic winds can be
suppressed in the cluster center while ram-pressure stripping is most
efficient there due to the fact that the ICM density as well as the
galaxies velocities are larger in the cluster center
(\citealt{2006A&A...447..827K}). \\ X-ray spectra are the only measure
for the metallicity of the ICM.  The metallicity is derived mainly by
measuring the equivalent width of the iron line once the continuum
(almost entirely given by thermal bremsstrahlung) is known. With the
first generation of satellites it was just possible to determine the
radial metallicity profiles (e.g. \citealt{1997ApJ...481L..63M},
\citealt{2004A&A...419....7D}). With deep observations of bright
clusters of galaxies by \chandra and \xmm satellites it is now
possible to extract metallicities in certain regions of a galaxy
cluster and construct X-ray weighted metallicity maps
(\citealt{2002MNRAS.337...71S}; \citealt{2004MNRAS.349..952S};
\citealt{2005A&A...432..809D}; \citealt{2005MNRAS.357.1134O};
\citealt{2005A&A...444..673S}; \citealt{2006A&A...449..475W};
\citealt{2006MNRAS.371.1483S}; \citealt{2006PASJ...58..695H};
\citealt{2009A&A...493..409S}; \citealt{2009A&A...508..191L}). \\ In
this paper we present the results of the analysis of a sample of 5
cool core clusters (Centaurus, Hydra A, S\'ersic 159-03, A496 and
A2029) observed with \xmm.  Our first goal is to show that even in
relaxed clusters the distribution of metals shows a lot of
inhomogeneities that cause an underestimation of the metal
mass. \\ The paper is structured as follows: in Sect. 2 we present the
data sets and data reduction techniques; in Sect. 3 we present
spatially resolved measurements of metals abundances; in Sect. 4 we
present the metallicity and temperature maps; in Sect. 5 we determine
the total, gas and metal mass of the clusters. A summary of our
conclusion is given in Sect 6.  Throughout the paper we assume
H$_{0}$=70 km s$^{-1}$ Mpc$^{-1}$, $\Omega_{\Lambda}$=0.73 and
$\Omega_{M}$=0.27. The elemental abundances presented in this paper
are given relative to the solar abundances from
\cite{1989GeCoA..53..197A}. We plot and tabulate values with error
quoted at the 90$\%$ level of confidence.

%__________________________________________________________________

\section{X-ray observations and analysis}

\subsection{Sample selection}
Our first aim is to derive the metallicity maps of the cluster. To
determine metallicities in many different regions of a cluster a lot
of photons are required. Therefore we selected clusters observed with
sufficiently long exposure from the \xmm archive. We found 5 clusters
suitable for our study: Centaurus, A496, S\'ersic 159-03, Hydra A and
A2029. We did not take the Perseus Cluster into account,
  because in literature there is already a detailed metallicity map
  obtained using Chandra data \citep{2005MNRAS.360..133S}. We decided
  to include Centaurus, because \cite{2006MNRAS.371.1483S} derived
  abundance maps for several elements but only for the inner part of
  the cluster ($\sim$80 kpc.)

\subsection{Data reduction}
Observation data files (ODFs) were retrieved from the XMM archive and
reprocessed with the \xmm Science Analysis System (SAS) v7.1.0. We
used tasks $emchain$ and $epchain$ to generate calibrated event files
from raw data. Throughout this analysis single pixel events for the pn
data (PATTERN 0) are selected, while for the MOS data sets the
PATTERNs 0-12 are used. In addition, for all cameras events next to
CCD edges and next to bad pixels were excluded (FLAG==0). \\ The data
were cleaned for periods of high background due to the soft proton
solar flares using a two stage filtering process. We first accumulated
in 100 s bins the light curve in the [10-12] keV band for MOS and
[12-14] keV for pn, where the emission is dominated by the
particle-induced background, and exclude all the intervals of
  exposure time having a count rate that deviated by more than 3$\sigma$
  from the mean (see \citealt{2002A&A...394..375P}) for 
  details. After filtering using the good time intervals from this
screening, the event lists was then re-filtered in a second pass as a
safety check for possible flares with soft spectra
(\citealt{2005ApJ...629..172N}; \citealt{2005A&A...443..721P}). In
this case light curves were made with 10 s bins in the full [0.3-10]
keV band. The resulting exposure times after cleaning are listed in
Table \ref{exptime}. \\ For the background subtraction we used a
combination of blank-sky maps and closed-filter observations as done
by \cite{2009A&A...493..409S}. From deep sky observations collected
with XMM, we selected the data with the most similar background for
each cluster. Both the blank-sky and the closed filter events were
selected by applying the same PATTERN selection, vignetting
correction, flare rejection criteria and point source removal used for
the observation events. In addition we transformed the coordinates of
the background files such that they were the same as for the
associated cluster data set. We calculated the count rates in the hard
energy band (10-12 keV for MOS and 12-14 keV for pn) outside of field
of view (OOFOV) for each observation, blank sky maps and closed filter
observations. For each detector we added to the corresponding blank
sky background set a fraction of the closed filter observation in
order to compensate for the difference between the OOFOV hard-band
count rate in the observation and in the blank-sky data. \\ To correct
for the vignetting effect, we used the photon weighting method
\citep{2001A&A...365L..80A}. The weight coefficients were computing by
applying the SAS task $evigweight$ to each event file. Point sources
were detected using the task $ewavelet$ in the energy band [0.3-10]
keV and checked by eye on images generated for each detector. We
produced a list of selected point sources from all available detectors
and the events in the corresponding regions were removed from both the
blank field and the observation data set.
\begin{figure*}[!t]
\hbox{ 
\epsfig{figure=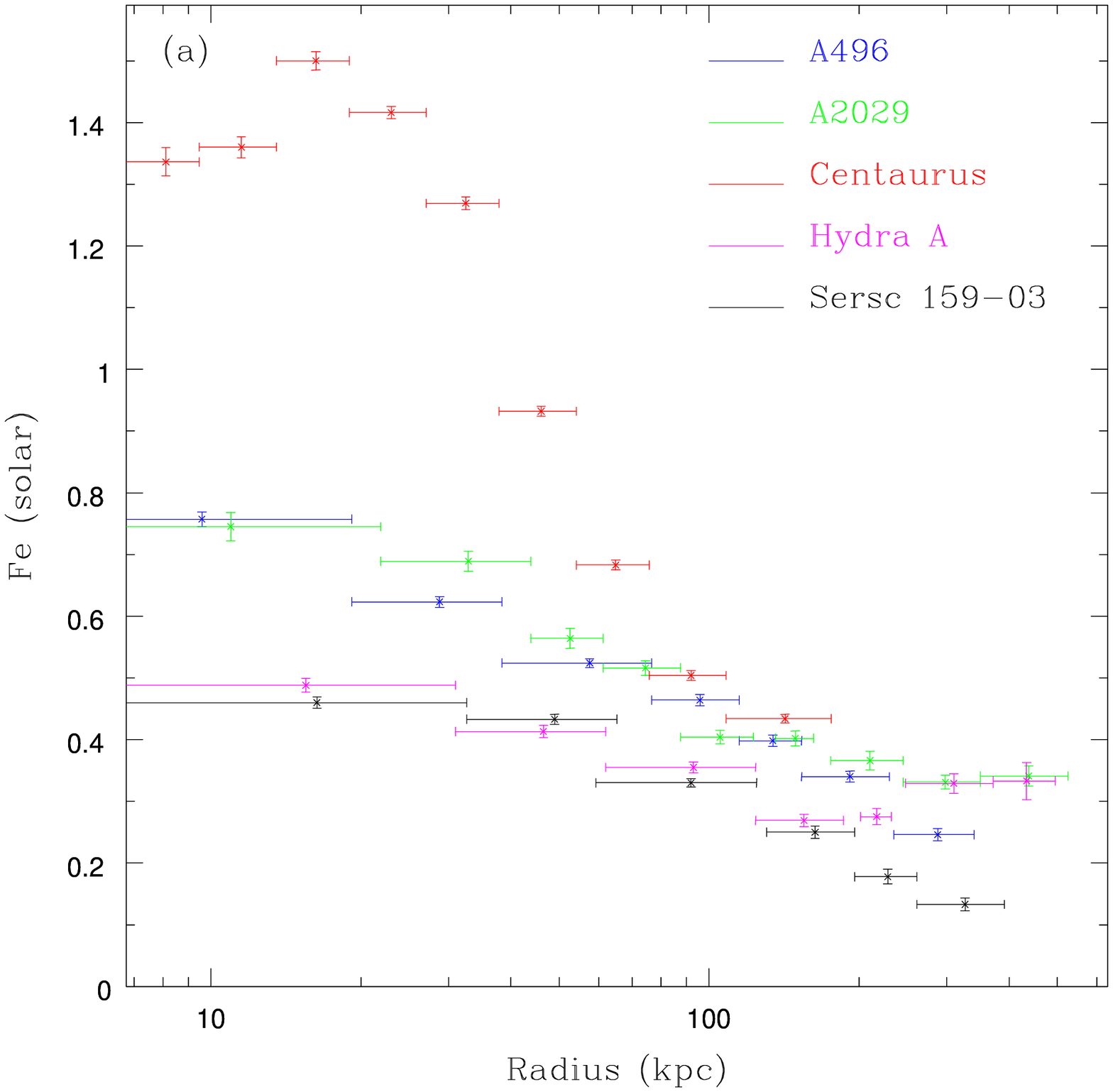,width=0.33\textwidth}
\epsfig{figure=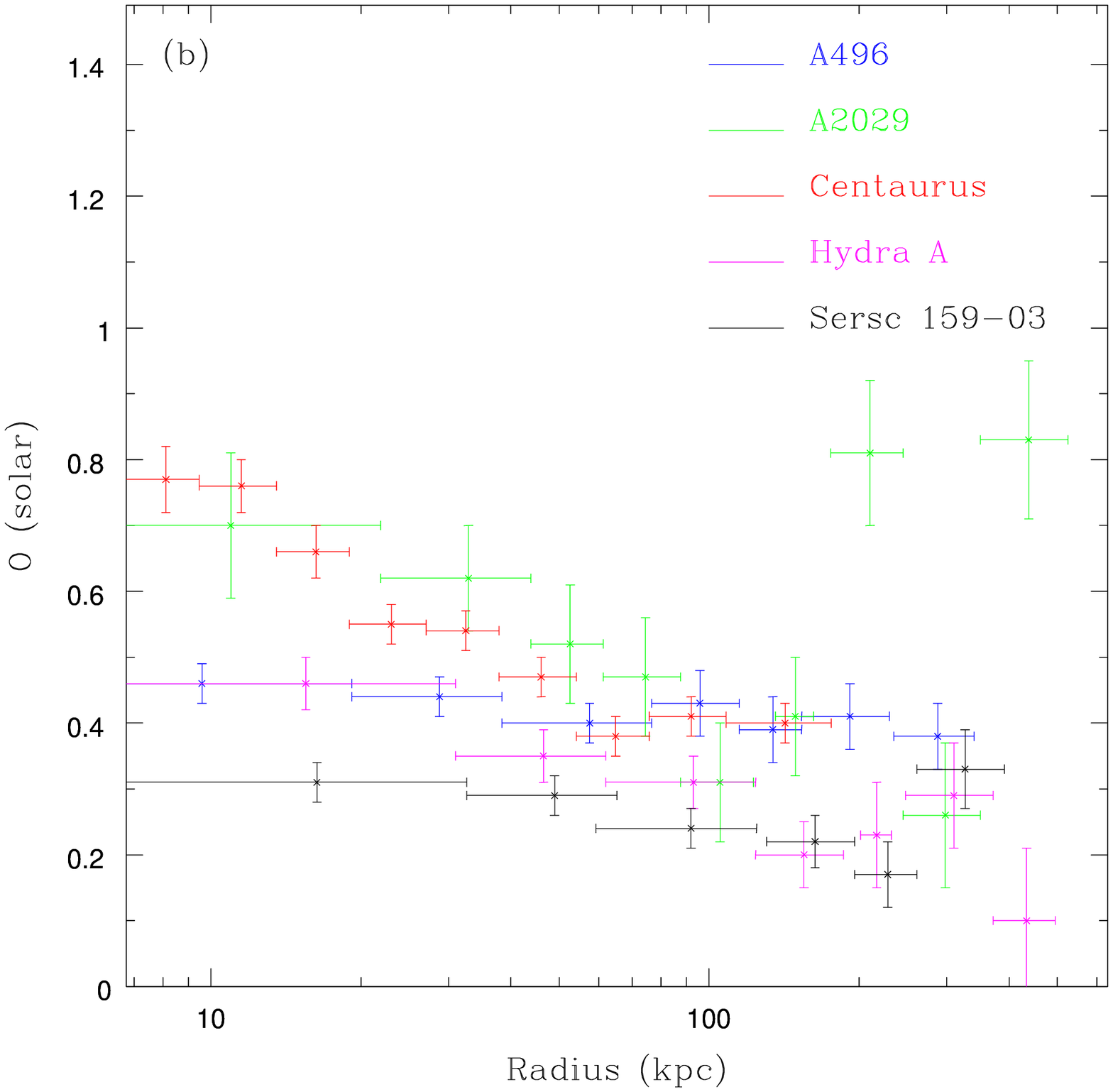,width=0.33\textwidth} 
\epsfig{figure=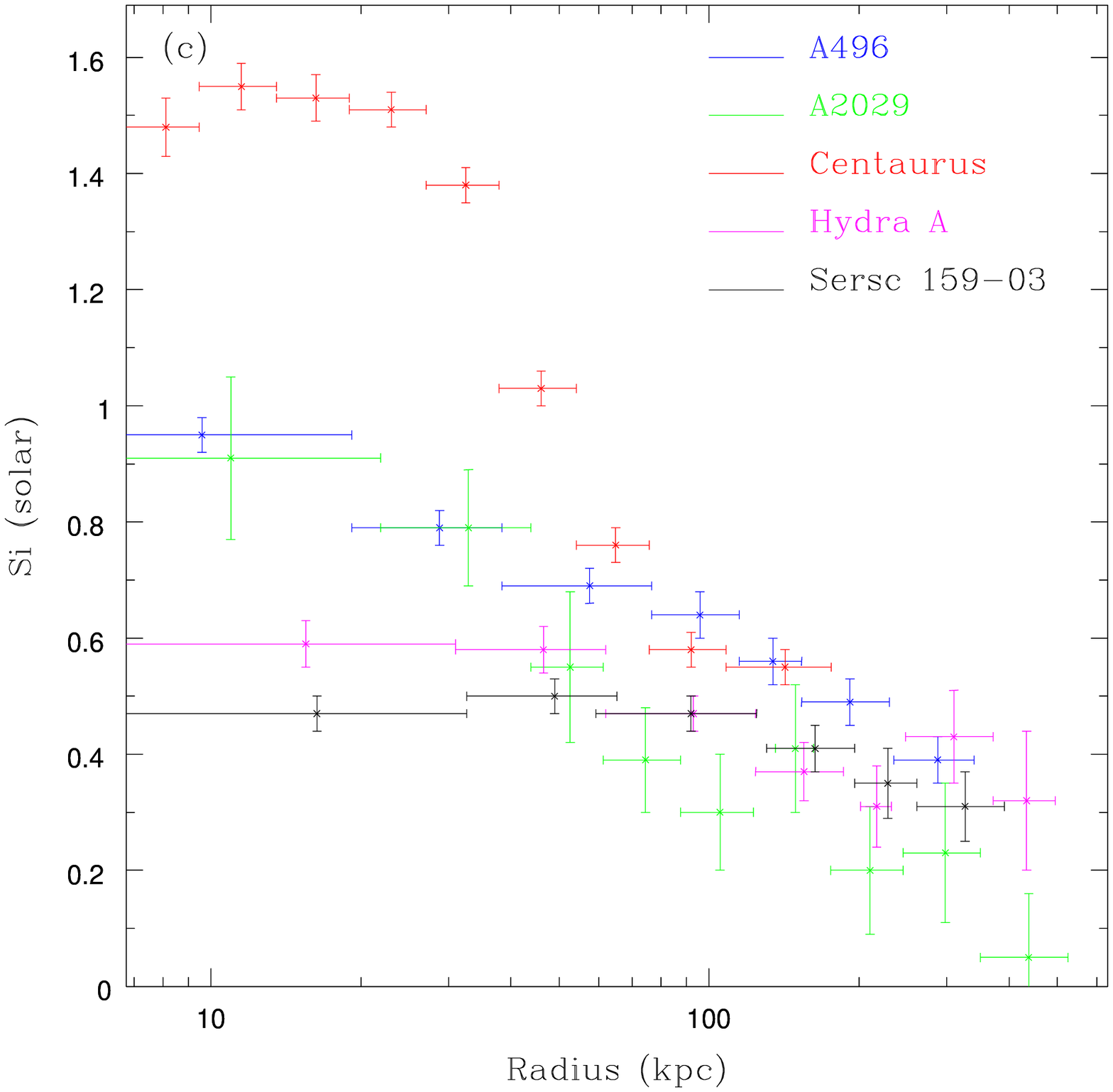,width=0.33\textwidth}
} 
\hbox{
\epsfig{figure=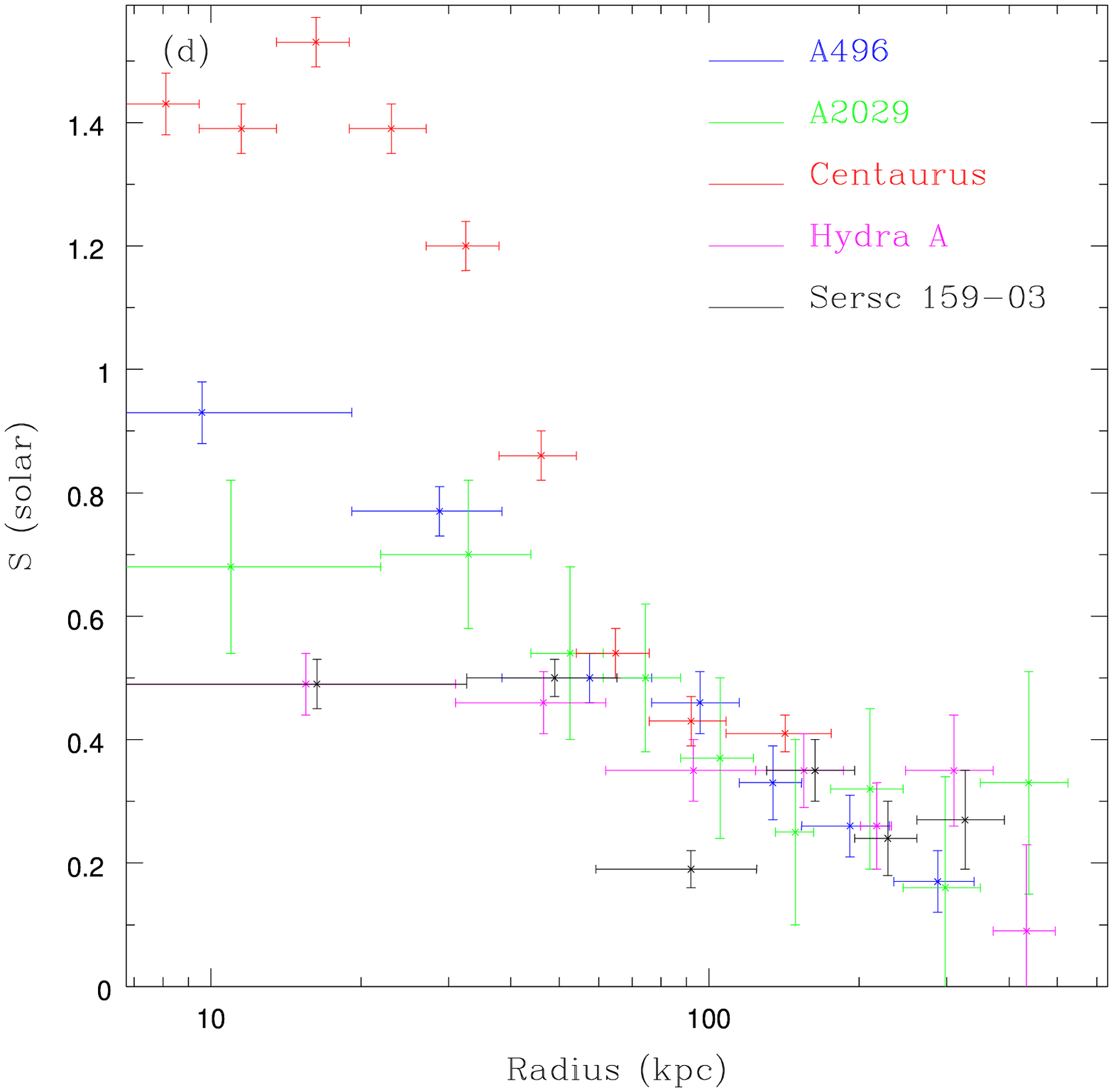,width=0.33\textwidth}
\epsfig{figure=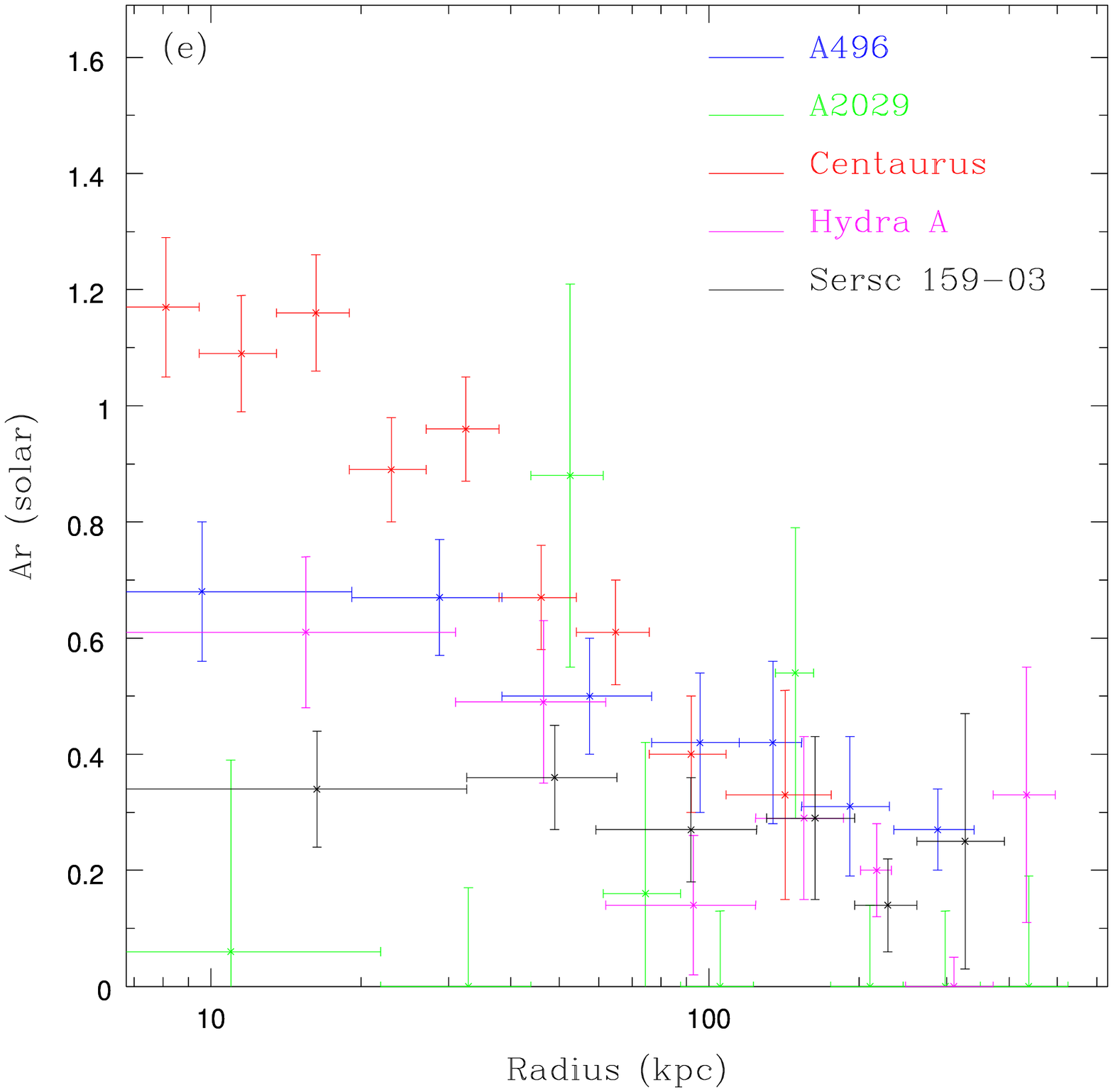,width=0.33\textwidth} 
}
\vspace*{-0.5ex}
\caption{Radial profiles of the (a) iron, (b)
  oxygen, (c) silicon, (d) sulfur and (e) argon in solar units,
  derived from the EPIC spectra. }
\label{fig:profiles} 
\end{figure*}
\subsection{Spectral analysis}
All the spectral fits were performed with the XSPEC package (version
12.5.0, \citealt{1996ASPC..101...17A}).  In order to model the
emission from a single (or multi) temperature plasma we fit the
spectra with an APEC (+APEC) model \citep{2001ApJ...556L..91S}
multiplied by the Galactic hydrogen column density, N$_H$, fixed at
the Galactic values \citep{1990ARA&A..28..215D} through the ${\it
  wabs}$ absorption model \citep{1983ApJ...270..119M}. To determine
the abundance profiles a VAPEC(+VAPEC) is used. We fit jointly MOS1,
MOS2 and pn spectra, enforcing the same normalization value for MOS
spectra and allowing the pn spectrum to have a separate
normalization. In the spectral fitting we used the 0.3-10 keV energy
range.  Because of the low number of counts and the high background at
large radii the spectra are fitted in the 0.3-9 keV band and 0.5-7.5
keV energy range for MOS and pn respectively. For the pn data we
excluded the energy above 7.5 keV in order to discard strong
instrumental lines around 8 keV. The redistribution and ancillary
files (RMF and ARF) were created with the SAS tasks ${\it rmfgen}$ and
${\it arfgen}$ for each camera and each region that we analyzed.

%__________________________________________________ One column table
   
\setlength{\extrarowheight}{0.1cm}
\begin{table} 
\caption{Cluster sample, sorted in order of increasing redshift. The exposure column indicates the net exposure time, after excluding time intervals of strong background flares. } \label{exptime}
    $$ 
         \centering
            \begin{array}{c c c c c c}
            \hline
            \noalign{\smallskip}
            Cluster & redshift & Orbit \ Rev. &      & exposure \ (ks) & \\
                    &          &            & MOS1 &   MOS2   & pn \\
            \noalign{\smallskip}
            \hline
            \noalign{\smallskip}
            Centaurus           & 0.0114 & 0379 & 28.0  & 35.8  & 31.5    \\
                                &        & 1213 & 107.5 & 106.5 & 80.6    \\
            A496                & 0.0329 & 1405 & 58.8  & 58.9  & 41.8   \\
                                &        & 1501 & 61.2  & 60.0  & 46.4    \\
            Hydra  A             & 0.0539 & 1359 & 81.6  & 82.4  & 55.1    \\
            S$\'e$rsic \ 159-03 & 0.0564 & 0540 & 92.9  & 94.2  & 73.9    \\         
            A2029               & 0.0773 & 0496 & 12.1  & 12.1  & 9.2    \\
                                &        & 1576 & 34.3  & 37.6  & 19.5    \\
                                &        & 1577 & 41.7  & 43.8  & 27.3     \\ 
                                &        & 1578 & 33.5  & 26.4  & 18.6    \\ 
                        \noalign{\smallskip}
            \hline
            \end{array}         
    $$ 
   \end{table}

\section{Abundances Profiles}
\begin{figure*}[!t]
\hbox{ \epsfig{figure=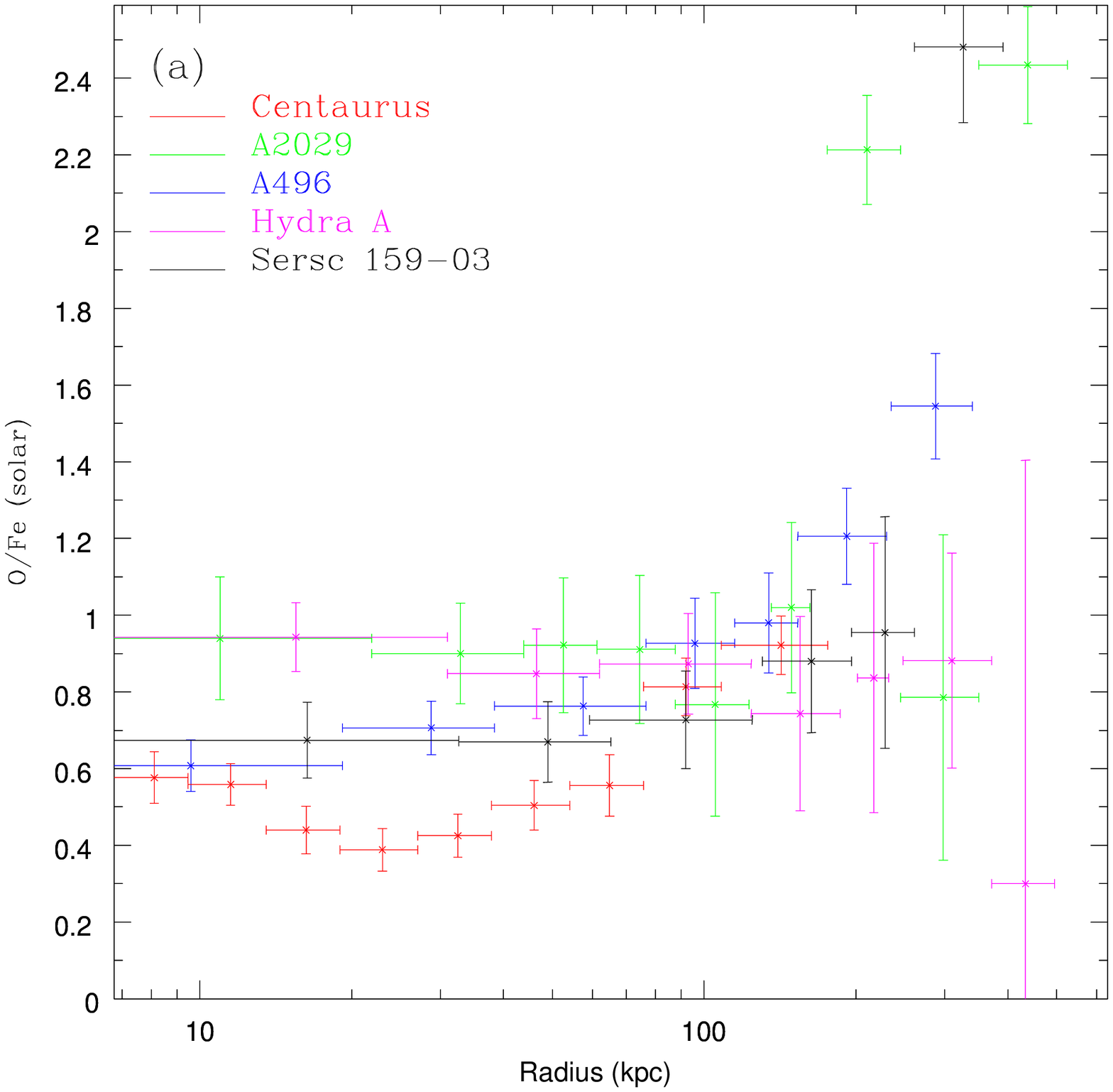,width=0.33\textwidth}
  \epsfig{figure=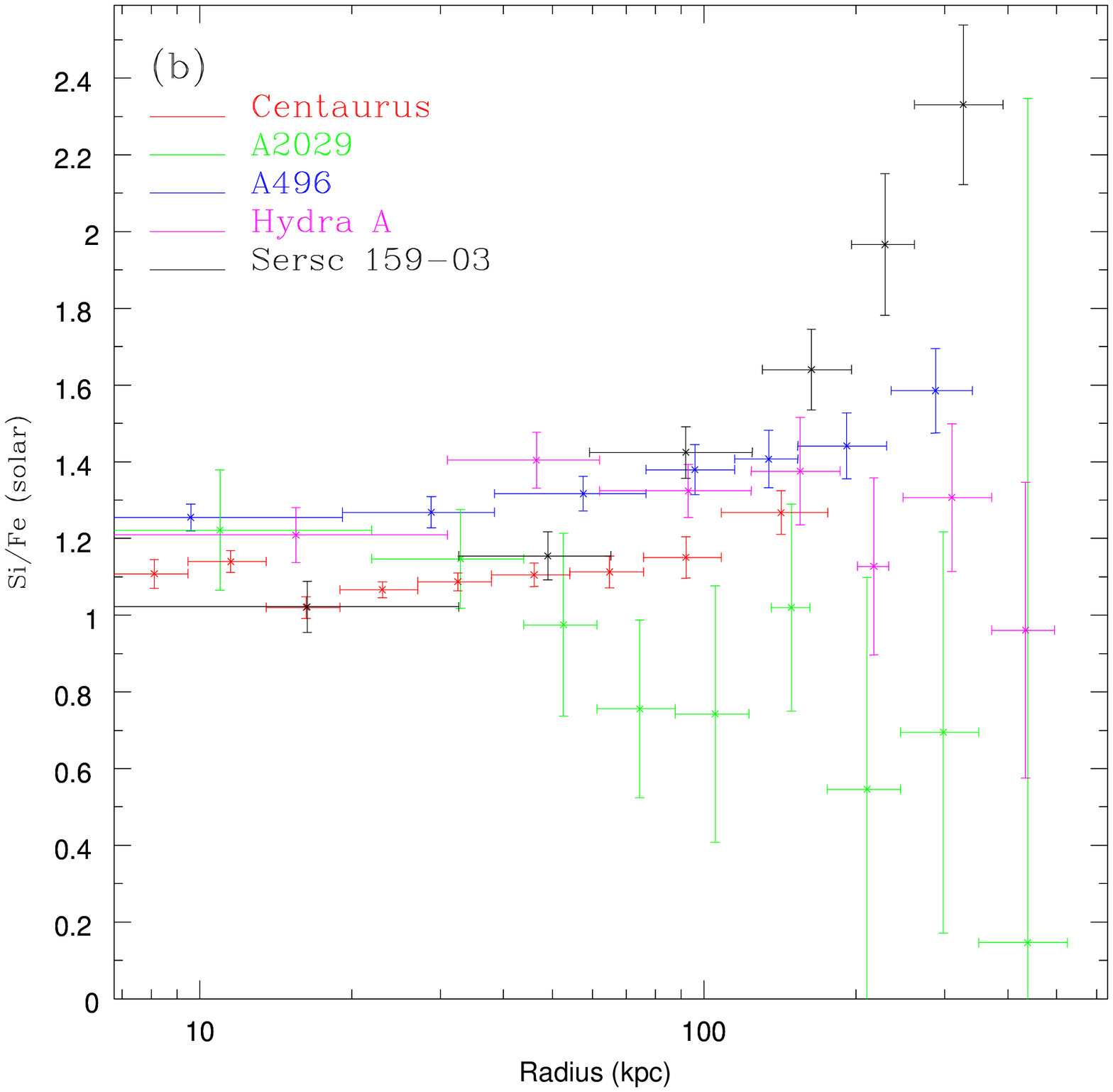,width=0.33\textwidth}
  \epsfig{figure=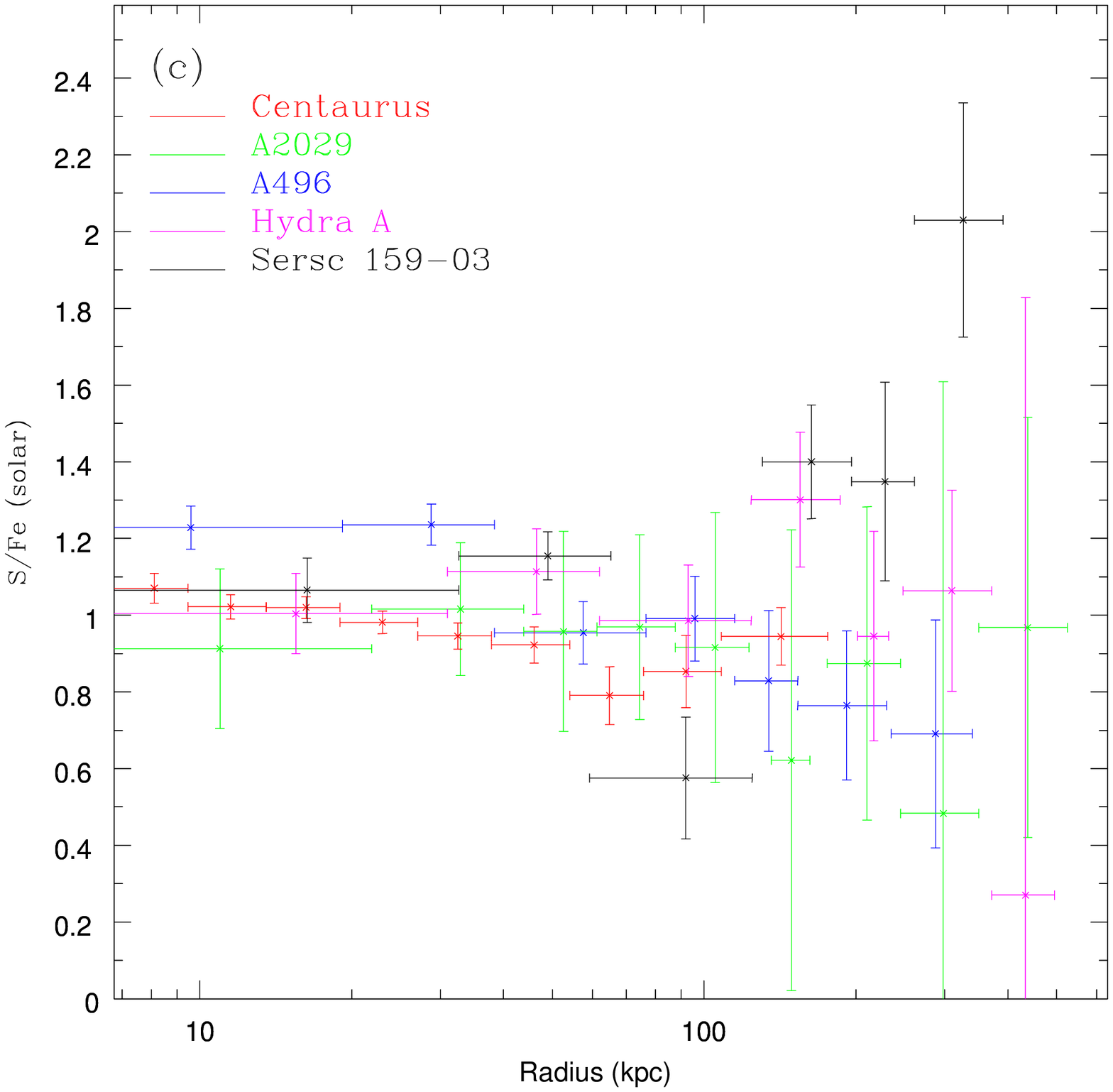,width=0.33\textwidth} } 
\vspace*{-0.5ex}
\caption{Radial profiles of the (a) O/Fe , (b) Si/Fe and (c)
  S/Fe ratio in solar units. }
\label{fig:xtofe} 
\end{figure*}

In order to know how the ICM has been enriched, we need to measure the
amount and distribution of metals in the ICM. Since Ar and Fe are
synthetized mainly in SNe Ia, O in core-collapse supernovae (SNCC), S
and Si in both SN Ia and SNCC, the measure of these elements can give
hints on the past enrichment process in the ICM by supernovae. \\ For
each cluster we extracted the spectra from several annular
regions. Metal abundance profiles are determined for O, Si, S, Ar and
Fe.  We fitted the data with the procedure presented in
\cite{2009A&A...508..191L} to avoid the degeneracy of the parameters:
(1) we fitted the data with an absorbed APEC model in the 0.4-7 keV
band to obtain the temperature (metallicity and normalization are
considered free parameters); (2) we fixed the temperature and use a
VAPEC model in the same energy band to determine the iron abundance
(O, Mg, Si, S, Ar, Ni were left free. The other elements were fixed to
the solar value); (3) we kept temperature and iron fixed to measure
the oxygen abundance in the 0.4-1.5 keV band; (4) we fix the values of
temperature, iron and oxygen to estimate the silicon, sulfur and argon
abundances in the 1.5-5 keV band. Fixing to the solar value the
abundances of elements, which cannot be significantly detected, may
introduce a bias in the abundance measurements of other elements if
the average cluster abundance is not solar. Thus, as a safety check we
fitted the spectra with C, N and Ne fixed at 0.3. Apart from Centaurus
the resulting abundance of the interested elements (O, Si, S, Fe) do
not change significantly. Furthermore, we note that in general, when
we examined all abundances to be free in the fits, the resulting
parameters did not change within the statistical errors.\\ Since there
are still cross-calibration issues at energies below $\sim 2 keV$ we
fit the MOS and pn spectra separately to investigate how robust the
derived values are. In general the values appear to be consistent
within the errors in the two instruments although for some annuli the
abundances of several elements are clearly not. In particular O shows
a strong discrepancy for Centaurus, A2029 and Hydra A up to 0.5 arcmin
while Si is not consistent for Sersic and Hydra A for the annulus
between 0.5 and 1 arcmin. The systematic difference between MOS and pn
is 3$\%$ for Fe, 17$\%$ for O and 11$\%$ for Si.  \\ In general the
single temperature model (1T) provided a good fit to the spectra.
However, in the inner region of all the clusters a 2T model
significantly improved the fit with respect to a single temperature
model as already found by \cite{2006MNRAS.371.1483S} and
\cite{2007A&A...462..953M} for Centaurus, by
\cite{2009A&A...493..409S} for Hydra A and by
\cite{2006A&A...452..397D} for S\'ersic 159-03. The Fe abundances
changed significantly as compared to the single temperature fit (the
so-called Fe-bias, see for example \citealt{2000MNRAS.311..176B} and
\citealt{2001A&A...375L..14M}). For these regions we use the
abundances derived from this 2T model for the discussion below.
\\ The abundance value for Si, S, Ar and Fe and their radial variation
look quite similar to each other. They commonly decline to about 1/4
of the central value in the outermost annulus (see
Fig. \ref{fig:profiles}).  The error on the Si, S, and Ar abundance
determination is larger than on iron. Nevertheless, their profiles
indicate a central increase similar to the iron distribution.\\
\begin{table*}[!ht]
\begin{center}
\caption{Fit results for spatially resolved EPIC spectra. For each
  cluster we show also the results obtained fitting the spectra of a
  radius encompassing an overdensity of $\sim4500$.  }
\label{abundances}
\begin{tabular}{cccccccccc}
\hline
Target	& radius & kT1 & kT2 & Fe & O/Fe & Si/Fe & S/Fe & Ar/Fe & $\chi^2/d.o.f.$ \\
	& arcmin & keV & keV &    &       &      &       &  \\ 
\hline\hline
Centaurus & 0-0.25   & 1.763$\pm$0.015 & 0.778$\pm$0.002 & 1.190$\pm$0.002 & 0.48$\pm$0.03 & 0.97$\pm$0.03 & 1.01$\pm$0.05 & 0.77$\pm$0.12 & 1492/984  \\
          & 0.25-0.5 & 1.853$\pm$0.008 & 0.846$\pm$0.003 & 1.276$\pm$0.017 & 0.49$\pm$0.04 & 1.03$\pm$0.03 & 0.98$\pm$0.04 & 1.05$\pm$0.10 & 2149/1290 \\
          & 0.5-0.7  & 1.969$\pm$0.006 & 1.001$\pm$0.010 & 1.336$\pm$0.023 & 0.58$\pm$0.04 & 1.11$\pm$0.05 & 1.07$\pm$0.06 & 0.88$\pm$0.10 & 1917/1303 \\
          & 0.7-1    & 2.120$\pm$0.010 & 1.078$\pm$0.009 & 1.360$\pm$0.017 & 0.56$\pm$0.04 & 1.14$\pm$0.04 & 1.02$\pm$0.04 & 0.80$\pm$0.09 & 2390/1510 \\
          & 1-1.4    & 2.462$\pm$0.009 & 1.275$\pm$0.010 & 1.500$\pm$0.015 & 0.44$\pm$0.03 & 1.02$\pm$0.04 & 1.02$\pm$0.04 & 0.77$\pm$0.08 & 2652/1743 \\
          & 1.4-2    & 2.755$\pm$0.008 &                 & 1.416$\pm$0.010 & 0.39$\pm$0.02 & 1.07$\pm$0.03 & 0.98$\pm$0.04 & 0.69$\pm$0.06 & 3168/1963 \\
          & 2-2.8    & 2.998$\pm$0.008 &                 & 1.269$\pm$0.010 & 0.42$\pm$0.03 & 1.09$\pm$0.03 & 0.95$\pm$0.03 & 0.76$\pm$0.07 & 3722/3110 \\
          & 2.8-4    & 3.368$\pm$0.008 &                 & 0.932$\pm$0.008 & 0.50$\pm$0.04 & 1.10$\pm$0.05 & 0.92$\pm$0.05 & 0.72$\pm$0.10 & 3363/2258 \\
          & 4-5.6    & 3.685$\pm$0.012 &                 & 0.683$\pm$0.008 & 0.56$\pm$0.05 & 1.11$\pm$0.06 & 0.79$\pm$0.07 & 0.89$\pm$0.15 & 3222/2448 \\
          & 5.6-8    & 3.745$\pm$0.013 &                 & 0.504$\pm$0.008 & 0.81$\pm$0.08 & 1.15$\pm$0.08 & 0.85$\pm$0.10 & 0.79$\pm$0.22 & 3159/2386 \\
          & 8-13     & 3.800$\pm$0.016 &                 & 0.434$\pm$0.010 & 0.92$\pm$0.09 & 1.27$\pm$0.09 & 0.94$\pm$0.09 & 0.76$\pm$0.43 & 3503/2436 \\
          & 0-13     & 3.773$\pm$0.016 & 1.662$\pm$0.007 & 0.649$\pm$0.010 & 0.64$\pm$0.04 & 1.30$\pm$0.07 & 1.13$\pm$0.08 & 1.01$\pm$0.15 & 2590/2051 \\
\hline
A496      & 0-0.5    & 2.903$\pm$0.022 & 1.332$\pm$0.012 & 0.757$\pm$0.012 & 0.61$\pm$0.05 & 1.25$\pm$0.06 & 1.23$\pm$0.08 & 0.90$\pm$0.17 & 2115/1688 \\
          & 0.5-1    & 3.675$\pm$0.071 & 1.710$\pm$0.027 & 0.626$\pm$0.009 & 0.71$\pm$0.05 & 1.27$\pm$0.06 & 1.24$\pm$0.07 & 1.07$\pm$0.18 & 2347/1951 \\
          & 1-2      & 4.462$\pm$0.140 & 2.118$\pm$0.018 & 0.524$\pm$0.007 & 0.76$\pm$0.07 & 1.32$\pm$0.07 & 0.95$\pm$0.09 & 0.95$\pm$0.21 & 2897/2263 \\
          & 2-3      & 3.944$\pm$0.018 &                 & 0.464$\pm$0.009 & 0.93$\pm$0.13 & 1.37$\pm$0.12 & 0.99$\pm$0.13 & 0.91$\pm$0.28 & 2734/2204 \\
          & 3-4      & 4.060$\pm$0.022 &                 & 0.398$\pm$0.009 & 1.08$\pm$0.15 & 1.40$\pm$0.13 & 0.83$\pm$0.17 & 1.05$\pm$0.39 & 2491/2114 \\
          & 4-6      & 4.100$\pm$0.025 &                 & 0.340$\pm$0.009 & 1.20$\pm$0.18 & 1.44$\pm$0.20 & 0.76$\pm$0.15 & 0.91$\pm$0.39 & 2674/2095 \\
          & 6-9      & 3.981$\pm$0.025 &                 & 0.246$\pm$0.010 & 1.54$\pm$0.28 & 1.59$\pm$0.23 & 0.69$\pm$0.24 & 1.10$\pm$0.34 & 2700/2026 \\
          & 0-5      & 3.453$\pm$0.024 & 2.081$\pm$0.018 & 0.525$\pm$0.010 & 0.71$\pm$0.11 & 1.16$\pm$0.12 & 0.83$\pm$0.13 & 0.88$\pm$0.24 & 2760/2214 \\

\hline
S\'ersic\ 159-03 & 0-0.5 & 2.783$\pm$0.027 & 1.937$\pm$0.013 & 0.460$\pm$0.012 & 0.67$\pm$0.06 & 1.08$\pm$0.07 & 1.06$\pm$0.12 & 0.74$\pm$0.24 & 1631/1429 \\
                 & 0.5-1 & 3.103$\pm$0.022 & 2.192$\pm$0.026 & 0.433$\pm$0.008 & 0.67$\pm$0.08 & 1.15$\pm$0.10 & 1.15$\pm$0.10 & 0.83$\pm$0.23 & 1731/1432 \\
                 & 1-2   & 2.574$\pm$0.012 &                 & 0.330$\pm$0.006 & 0.74$\pm$0.09 & 1.42$\pm$0.11 & 0.59$\pm$0.12 & 0.83$\pm$0.28 & 1846/1587 \\
                 & 2-3   & 2.505$\pm$0.022 &                 & 0.250$\pm$0.010 & 0.88$\pm$0.20 & 1.64$\pm$0.23 & 1.40$\pm$0.27 & 1.16$\pm$0.63 & 1386/1221 \\
                 & 3-4   & 2.357$\pm$0.032 &                 & 0.178$\pm$0.011 & 0.94$\pm$0.18 & 1.97$\pm$0.19 & 1.35$\pm$0.32 & 0.77$\pm$0.54 & 1226/1007 \\
                 & 4-6   & 2.063$\pm$0.020 &                 & 0.133$\pm$0.010 & 3.00$\pm$0.20 & 2.33$\pm$0.29 & 2.07$\pm$0.47 & 1.90$\pm$1.60 & 1201/1012 \\
                 & 0-2.5 & 2.590$\pm$0.018 & 1.303$\pm$0.026 & 0.405$\pm$0.007 & 0.76$\pm$0.06 & 1.09$\pm$0.05 & 0.83$\pm$0.05 & 0.79$\pm$0.14 & 2587/2030 \\
\hline
Hydra A   & 0-0.5    & 5.064$\pm$0.015 & 2.232$\pm$0.077 & 0.488$\pm$0.009 & 0.94$\pm$0.11 & 1.23$\pm$0.09 & 1.00$\pm$0.13 & 1.25$\pm$0.30 & 1979/1730 \\
          & 0.5-1    & 5.354$\pm$0.015 & 2.423$\pm$0.245 & 0.413$\pm$0.010 & 0.85$\pm$0.12 & 1.40$\pm$0.10 & 1.11$\pm$0.15 & 1.15$\pm$0.36 & 2029/1796 \\
          & 1-2      & 5.027$\pm$0.022 & 2.165$\pm$0.230 & 0.355$\pm$0.009 & 0.88$\pm$0.13 & 1.33$\pm$0.15 & 1.00$\pm$0.17 & 0.39$\pm$0.81 & 2271/1906 \\
          & 2-3      & 5.077$\pm$0.020 & 2.267$\pm$0.153 & 0.269$\pm$0.010 & 0.74$\pm$0.19 & 1.37$\pm$0.20 & 1.31$\pm$0.24 & 1.08$\pm$0.47 & 1929/1761 \\
          & 3-4      & 3.879$\pm$0.035 &                 & 0.273$\pm$0.013 & 0.84$\pm$0.28 & 1.13$\pm$0.28 & 0.94$\pm$0.31 & 0.73$\pm$0.56 & 1790/1657 \\
          & 4-6      & 3.760$\pm$0.034 &                 & 0.329$\pm$0.019 & 0.88$\pm$0.31 & 1.31$\pm$0.40 & 1.06$\pm$0.36 & 0.00$\pm$0.16 & 2076/1789 \\
          & 6-8      & 3.706$\pm$0.090 &                 & 0.333$\pm$0.020 & 0.31$\pm$0.37 & 0.97$\pm$0.45 & 0.27$\pm$0.27 & 0.99$\pm$0.77 & 2132/1724 \\
          & 0-3      & 5.090$\pm$0.127 & 2.183$\pm$0.039 & 0.370$\pm$0.008 & 0.66$\pm$0.06 & 1.21$\pm$0.07 & 1.02$\pm$0.07 & 0.66$\pm$0.16 & 3006/2505 \\
\hline
A2029     & 0-0.25   & 7.715$\pm$0.040 & 2.891$\pm$0.127 & 0.744$\pm$0.013 & 0.93$\pm$0.11 & 1.22$\pm$0.18 & 0.95$\pm$0.20 & 0.06$\pm$0.33 & 1750/1657 \\
          & 0.25-0.5 & 8.415$\pm$0.041 & 3.394$\pm$0.142 & 0.689$\pm$0.013 & 0.90$\pm$0.12 & 1.15$\pm$0.16 & 1.01$\pm$0.21 & 0.00$\pm$0.17 & 2461/2112 \\
          & 0.5-0.7  & 8.190$\pm$0.047 & 3.214$\pm$0.167 & 0.568$\pm$0.016 & 0.92$\pm$0.17 & 0.98$\pm$0.20 & 0.99$\pm$0.29 & 1.46$\pm$0.65 & 2288/2054 \\
          & 0.7-1    & 7.886$\pm$0.047 & 2,751$\pm$0.182 & 0.516$\pm$0.012 & 0.92$\pm$0.18 & 0.76$\pm$0.21 & 0.97$\pm$0.30 & 0.31$\pm$0.55 & 2467/2220 \\
          & 1-1.4    & 8.191$\pm$0.135 & 2.068$\pm$0.140 & 0.402$\pm$0.011 & 0.78$\pm$0.24 & 0.74$\pm$0.28 & 0.95$\pm$0.36 & 0.00$\pm$0.26 & 2508/2303 \\
          & 1.4-2    & 8.415$\pm$0.065 & 2.435$\pm$0.187 & 0.400$\pm$0.012 & 1.02$\pm$0.24 & 1.05$\pm$0.29 & 0.65$\pm$0.36 & 1.34$\pm$0.82 & 2571/2377 \\
          & 2-2.8    & 7.388$\pm$0.060 &                 & 0.366$\pm$0.015 & 2.21$\pm$0.31 & 0.55$\pm$0.34 & 0.88$\pm$0.34 & 0.00$\pm$0.41 & 2673/2303 \\
          & 2.8-4    & 7.704$\pm$0.066 &                 & 0.331$\pm$0.011 & 0.79$\pm$0.37 & 0.76$\pm$0.40 & 0.48$\pm$0.36 & 0.00$\pm$0.16 & 2682/2299 \\
          & 4-6      & 8.215$\pm$0.085 &                 & 0.341$\pm$0.016 & 2.43$\pm$0.49 & 0.15$\pm$0.34 & 0.97$\pm$0.60 & 0.00$\pm$0.58 & 3286/2470 \\
          & 0-3      & 8.100$\pm$0.087 & 2.385$\pm$0.123 & 0.470$\pm$0.016 & 1.12$\pm$0.13 & 0.94$\pm$0.13 & 1.00$\pm$0.16 & 0.43$\pm$0.07 & 3490/2791 \\
	& & & & & & &  &   \\
\hline
\hline\\
\end{tabular}
\end{center}
\end{table*}
\noindent The O profile looks more complex compared with the other
elements. It shows several discontinuities, probably due to the fact
that O is strongly related to episodes of star formation, in fact the
O abundance seems to increase in the outer rings, where we expect
higher star formation. On the other hand, while in the central regions
the flux is high enough to get an accurate measurement, at large
radii, the oxygen abundance could be overestimated because the oxygen
in the galactic foreground emission starts to play an important
role. This effect is most apparent if the background is high with
respect to the source emission. Apart from Centaurus, in the outermost
bin of each cluster we found approximately a S/N of 4, thus the O
abundance should be used with caution there.\\
\noindent We computed abundance ratio of O, Si, S and Ar over Fe as a
function of the projected radius (see Table \ref{abundances}).  The
ratios Si/Fe and S/Fe are consistent with a constant values around 1-2
and 0.7-1.5 respectively, while the O/Fe ratio for the innermost
region is lower around 0.5-1 (see Fig. \ref{fig:xtofe}).  In addition
the O/Fe ratio suggests some increase with radius with the exception
of Hydra A for which both O/Fe and Si/Fe ratio seems to be
constant. If we do not consider the outermost bin the O/Fe show a
  slight increase with radius in agreement with the results obtained
  by \cite{2009A&A...493..409S}. \\ Using Suzaku
observations, \citep{2007PASJ...59..299S, 2008arXiv0810.3820S,
  2008arXiv0805.2771S, 2008PASJ...60S.333S},
\cite{2008PASJ...60S.317T} and \cite{2007PASJ...59S.327M} have
presented abundances of groups and clusters of galaxies. All systems
show very similar value of Si/Fe ratio, to be 1-1.5, in good agreement
with our results. \cite{2004A&A...420..135T} reported abundance ratios
for 19 clusters (among them A496, Hydra A, S\'ersic but with shorter
exposure time) studied with \xmm, and the mean Si/Fe was
$\sim$1.4. Their O/Fe, $\sim$0.6, is bit lower than our results. A
ratio of 1-2 for the Si/Fe ratio was found also by
\cite{2009A&A...508..565D} analyzing 26 clusters.
\cite{2006MNRAS.371.1483S} showed the abundances ratios for the
Centaurus cluster with \chandra and \xmm, and the radial abundance
ratios of O/Fe and Si/Fe to be 0.5-1 and 1-1.5 respectively, were
consistent with our results.\\ Therefore, in general our results are
in agreement with previous studies suggesting that cluster and groups
have passed the same metal enrichment process in the ICM.

\setlength{\extrarowheight}{0.1cm}
\begin{table}[!h] \label{table:snratio}
\caption{Number ratio of SNCC to SN Ia (N$_{II}$/N$_{Ia}$ and
  integrated number of SNCC computed using the abundance of O for
  each considered region. For each cluster we show also the results
  obtaining fitting the spectra of a radius encompassing an
  overdensity of $\sim$4500.}

            \begin{tabular}{c c | c c c c  }
            \hline
            \noalign{\smallskip}
                                &        &  \multicolumn{4}{c}{N$_{II}$/ N$_{Ia}$ }   \\
            Cluster             & region & O/Fe & Si/Fe & S/Fe & Ar/Fe   \\
            \noalign{\smallskip}
            \hline
            \noalign{\smallskip}
Centaurus & 0-0.25   &  1.09$^{+0.08}_{-0.08}$ & 0.77$^{+0.12}_{-0.12}$ & 0.95$^{+0.42}_{-0.34}$ & $<$1.67               \\
          & 0.25-0.5 &  1.11$^{+0.11}_{-0.11}$ & 1.02$^{+0.13}_{-0.14}$ & 0.74$^{+0.29}_{-0.25}$ & 6.25$^{+3.74}_{-2.40}$ \\
          & 0.5-0.7  &  1.35$^{+0.12}_{-0.10}$ & 1.37$^{+0.24}_{-0.23}$ & 1.40$^{+0.51}_{-0.46}$ & 2.50$^{+2.04}_{-1.54}$  \\
          & 0.7-1    &  1.30$^{+0.11}_{-0.11}$ & 1.52$^{+0.18}_{-0.20}$ & 1.03$^{+0.30}_{-0.29}$ & 1.35$^{+3.19}_{-0.39}$  \\
          & 1-1.4    &  0.98$^{+0.07}_{-0.08}$ & 0.97$^{+0.16}_{-0.16}$ & 1.03$^{+0.30}_{-0.29}$ & 0.96$^{+1.08}_{-0.84}$  \\
          & 1.4-2    &  0.85$^{+0.05}_{-0.05}$ & 1.19$^{+0.13}_{-0.13}$ & 0.74$^{+0.29}_{-0.25}$ & $<$0.63                \\
          & 2-2.8    &  0.93$^{+0.07}_{-0.08}$ & 1.28$^{+0.13}_{-0.14}$ & 0.53$^{+0.21}_{-0.21}$ & 0.86$^{+0.89}_{-0.75}$ \\
          & 2.8-4    &  1.14$^{+0.11}_{-0.11}$ & 1.32$^{+0.25}_{-0.22}$ & 0.34$^{+0.34}_{-0.31}$ & $<$1.61  \\
          & 4-5.6    &  1.30$^{+0.15}_{-0.14}$ & 1.37$^{+0.19}_{-0.21}$ & - & 2.63$^{+3.25}_{-1.99}$                            \\
          & 5.6-8    &  2.04$^{+0.29}_{-0.25}$ & 1.54$^{+0.42}_{-0.35}$ & $<$0.53 & $<$5.00                                 \\
          & 8-13     &  2.44$^{+0.34}_{-0.31}$ & 2.17$^{+0.53}_{-0.48}$ & $<$1.26 & $<$12.5                            \\
          & 0-13     &  1.54$^{+0.10}_{-0.13}$ & 2.33$^{+0.37}_{-0.37}$ & 1.92$^{+0.78}_{-0.67}$ & 5.00$^{+5.00}_{-2.73}$  \\
\hline
A496      & 0-0.5    &  1.35$^{+0.14}_{-0.13}$ & 1.69$^{+0.34}_{-0.32}$ & 1.37$^{+0.55}_{-0.56}$ & $<$2.49              \\
          & 0.5-1    &  1.75$^{+0.21}_{-0.17}$ & 2.04$^{+0.52}_{-0.43}$ & 3.12$^{+1.22}_{-0.90}$ & 12.5$^{+24.7}_{-4.50}$ \\
          & 1-2      &  1.85$^{+0.23}_{-0.21}$ & 2.44$^{+0.42}_{-0.40}$ & $<$0.54 & 3.03$^{+5.29}_{-2.81}$              \\
          & 2-3      &  2.38$^{+0.48}_{-0.46}$ & 2.70$^{+0.76}_{-0.88}$ & 1.25$^{+1.07}_{-0.85}$ & 3.03$^{+3.22}_{-2.07}$ \\
          & 3-4      &  3.03$^{+0.54}_{-0.59}$ & 2.94$^{+0.99}_{-0.81}$ & $<$1.27 & 2.94$^{+3.31}_{-2.15}$              \\
          & 4-6      &  3.45$^{+0.54}_{-0.51}$ & 3.12$^{+1.11}_{-0.91}$ & $<$2.69 & 4.50$^{+12.1}_{-3.80}$              \\
          & 6-9      &  5.55$^{+2.77}_{-2.11}$ & 4.35$^{+2.31}_{-1.85}$ & $<$1.10 & 7.70$^{+42.0}_{-6.30}$              \\
          & 0-5      &  1.72$^{+0.36}_{-0.31}$ & 1.61$^{+0.61}_{-0.55}$ & $<$0.60 & $<$8.33 \\

\hline
S\'ersic  & 0-0.5    & 1.61$^{+0.17}_{-0.16}$ & 1.14$^{+0.31}_{0.29}$ & 1.33$^{+1.11}_{0.0.87}$ & $<$2.75  \\
          & 0-5-1    & 1.61$^{+0.24}_{-0.22}$ & 1.56$^{+0.48}_{-0.46}$ & 2.13$^{+1.00}_{-0.86}$ & $<$4.91  \\
          & 1-2      & 2.17$^{+0.26}_{-0.32}$ & 2.70$^{+0.63}_{-0.58}$ & $<$0.88 & $<$5.03  \\
          & 2-3      & 2.27$^{+0.76}_{-0.63}$ & 4.54$^{+2.12}_{-2.42}$ & 5.26$^{+5.84}_{-3.26}$ & -  \\
          & 3-4      & 2.50$^{+0.73}_{-0.61}$ & 7.69$^{+3.42}_{1.80}$ & 4.35$^{+6.76}_{-3.24}$ & $<$24.0  \\
          & 4-6      & 25.0$^{+8.50}_{-11.0}$ & 14.3$^{+10.7}_{-5.10}$ & - & -  \\
          & 0-2.5    & 1.89$^{+0.18}_{-0.20}$ & 1.28$^{+0.23}_{-0.23}$ &  $<$0.19 & $<$3.33  \\
\hline
Hydra A   & 0-0.5    &  2.38$^{+0.32}_{-0.38}$ & 1.61$^{+0.51}_{-0.48}$ & $<$0.86 & 12.5$^{+10.0}_{-10.0}$ \\
          & 0.5-1    &  2.08$^{+0.48}_{-0.44}$ & 2.70$^{+0.63}_{-0.66}$ & 1.49$^{+1.28}_{-1.00}$ & -   \\
          & 1-2      &  2.27$^{+0.50}_{-0.42}$ & 2.50$^{+0.95}_{-0.80}$ & $<$1.41 & $<$19 \\
          & 2-3      &  1.82$^{+0.62}_{-0.55}$ & 2.70$^{+1.30}_{-1.06}$ & 3.84$^{+4.48}_{-2.44}$ & -   \\
          & 3-4      &  2.12$^{+0.65}_{-0.86}$ & 1.41$^{+1.71}_{-1.21}$  & $<$2.63 & $<$24.5  \\
          & 4-6      &  2.33$^{+0.90}_{-0.81}$ & 2.38$^{+1.62}_{-1.22}$  & - & -  \\
          & 6-8      &  $<$1.61 & $<$3.12 & - & $<$4.76   \\
          & 0-3      &  1.59$^{+0.16}_{-0.18}$ & 1.85$^{+0.37}_{-0.34}$ & 1.03$^{+0.56}_{-0.58}$ & $<$1.61  \\
\hline
A2029     & 0-0.25   &  2.44$^{+0.42}_{-0.36}$ & 1.89$^{+1.05}_{-0.82}$ & $<$2.13 & -   \\
          & 0.25-0.5 &  2.38$^{+0.40}_{-0.42}$ & 1.54$^{+0.84}_{-0.69}$ & $<$2.85 & -   \\
          & 0.5-0.7  &  2.44$^{+0.59}_{-0.59}$ & 0.81$^{+0.89}_{-0.74}$ & $<$3.57 & -   \\
          & 0.7-1    &  2.44$^{+0.68}_{-0.60}$ & $<$0.76 & $<$3.27 & $<$2.50   \\
          & 1-1.4    &  1.96$^{+0.82}_{-0.71}$ & $<$0.97 & $<$3.84 & -                \\
          & 1.4-2    &  2.78$^{+0.93}_{-0.82}$ & $<$2.55 & $<$0.95 & -   \\
          & 2-2.8    &  9.99$^{+4.28}_{-2.30}$ & $<$0.46 & $<$2.71 & -                \\
          & 2.8-4    &  1.88$^{+1.16}_{-0.90}$ & $<$1.61 & - & -                      \\
          & 4-6      &  12.5$^{+12.5}_{-4.80}$ & - & $<$8.32  & -                     \\
          & 0-3      &  3.12$^{+0.58}_{-0.58}$ & 0.66$^{+0.48}_{-0.46}$ & $<$2.22 & -  \\            
            \hline

                        \noalign{\smallskip}
            \hline
            \end{tabular}         
    
\end{table}

\begin{figure*}[!htb]
\hbox{ 
      \epsfig{figure=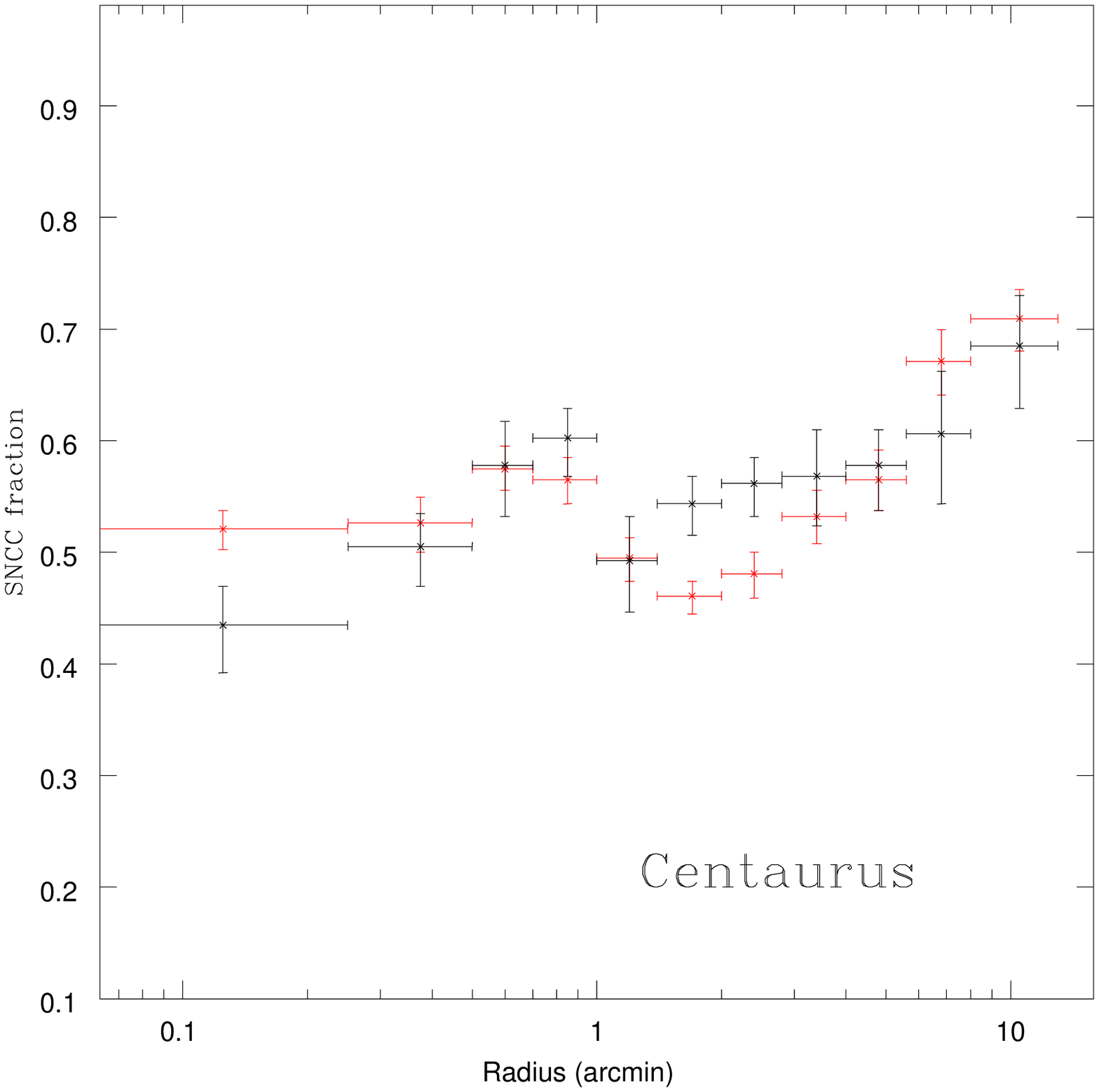,width=0.33\textwidth} 
      \epsfig{figure=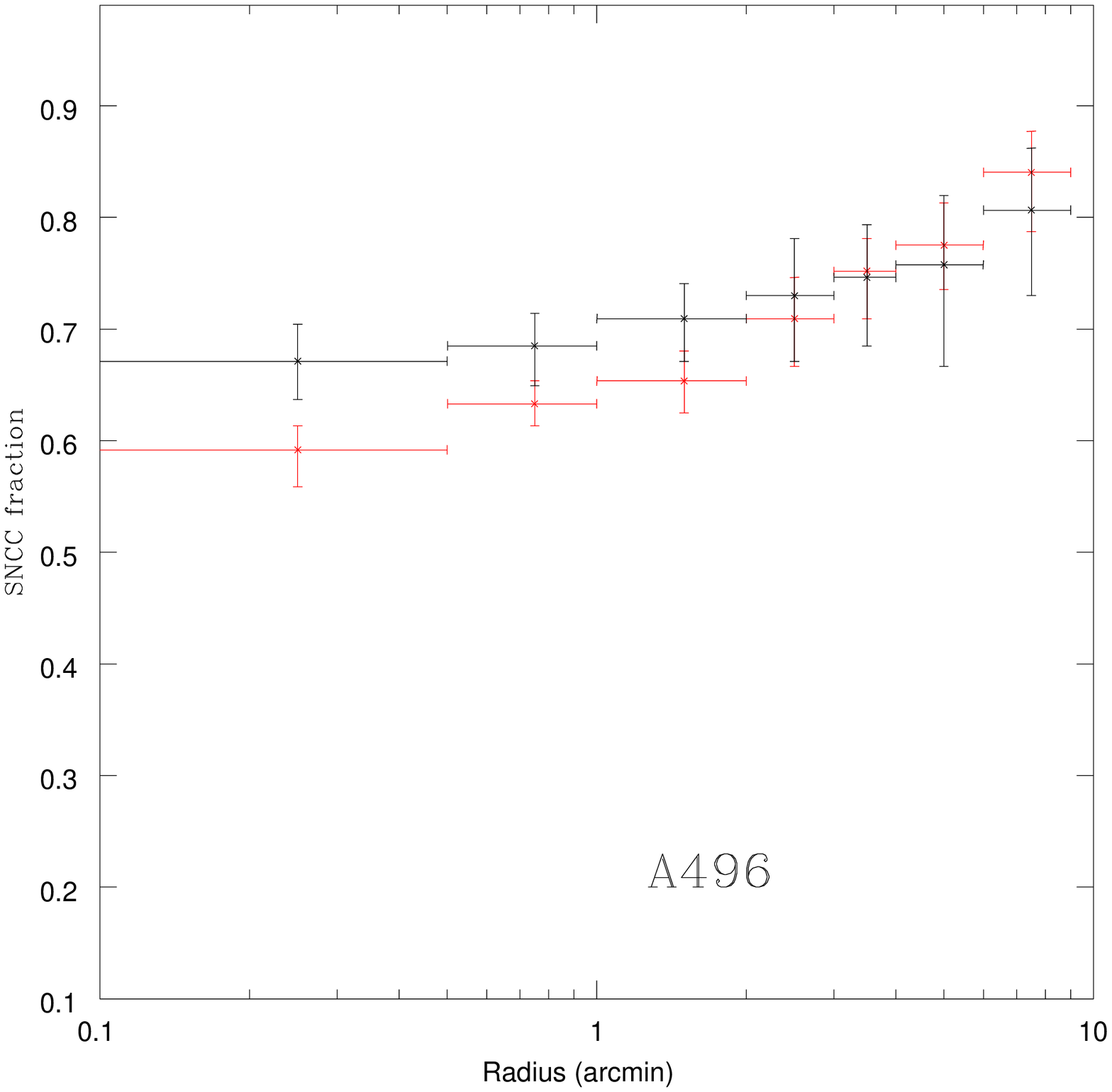,width=0.33\textwidth}  
      \epsfig{figure=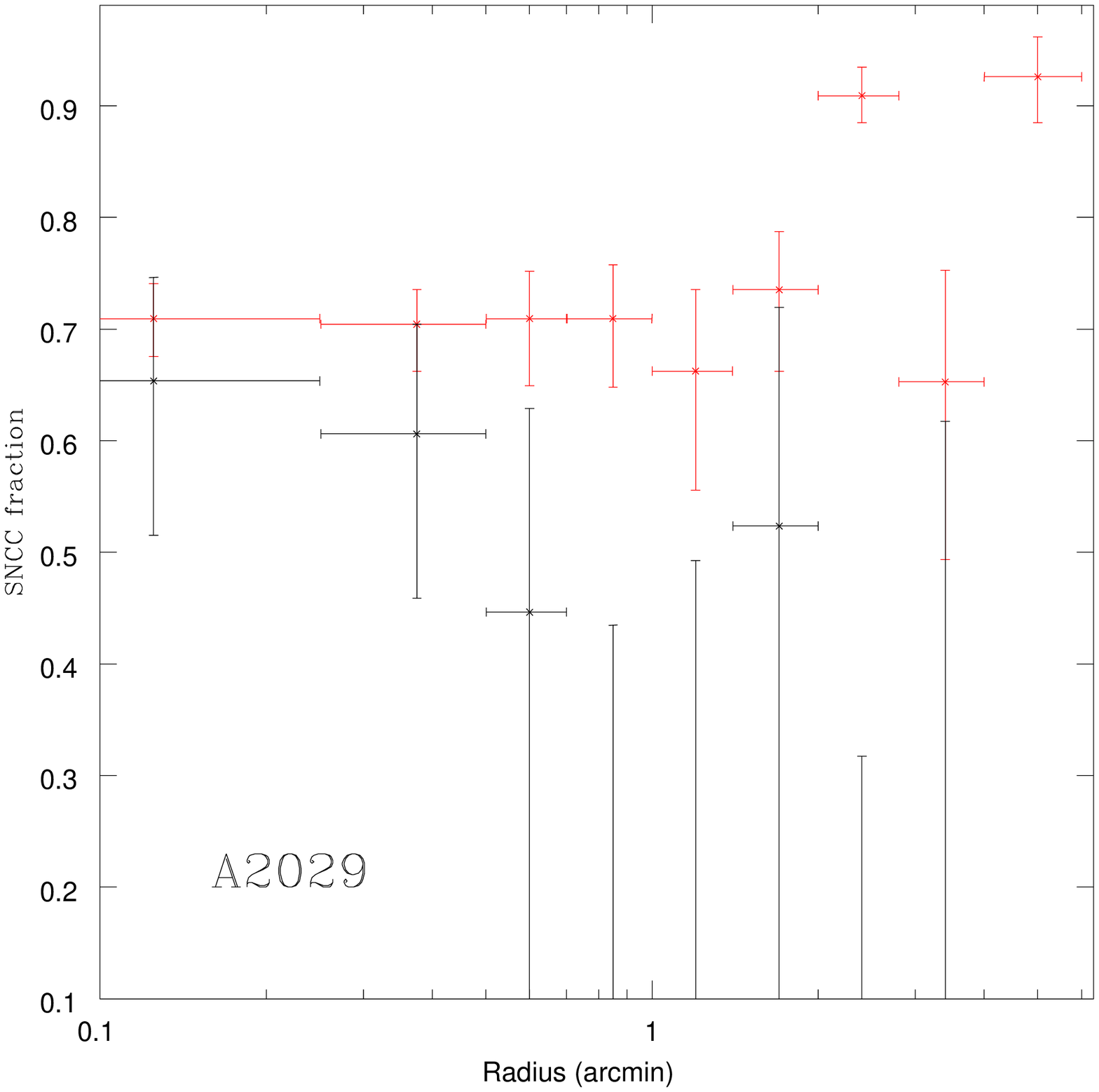,width=0.33\textwidth}  
}
\hbox{ 
      \epsfig{figure=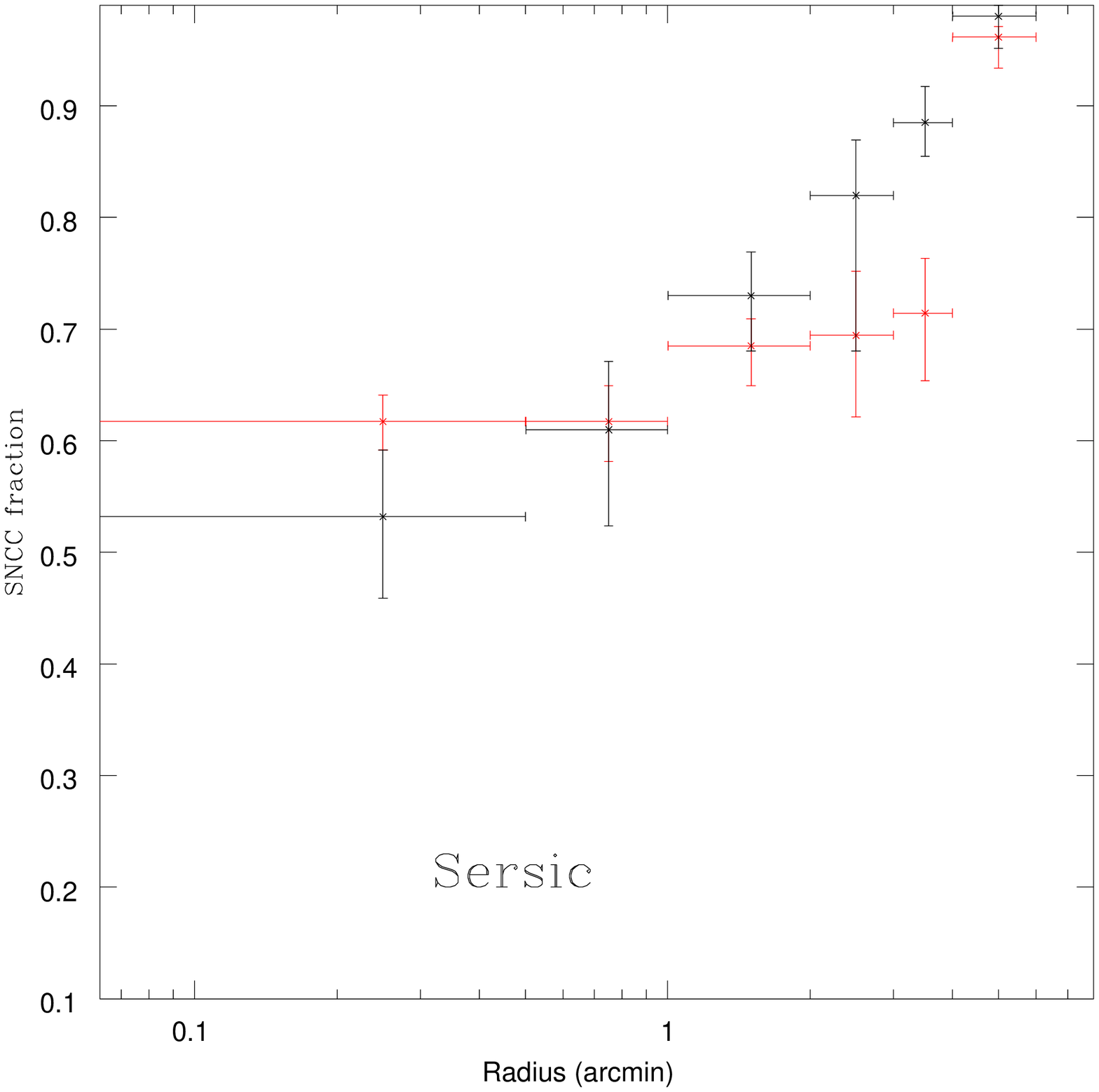,width=0.33\textwidth} 
      \epsfig{figure=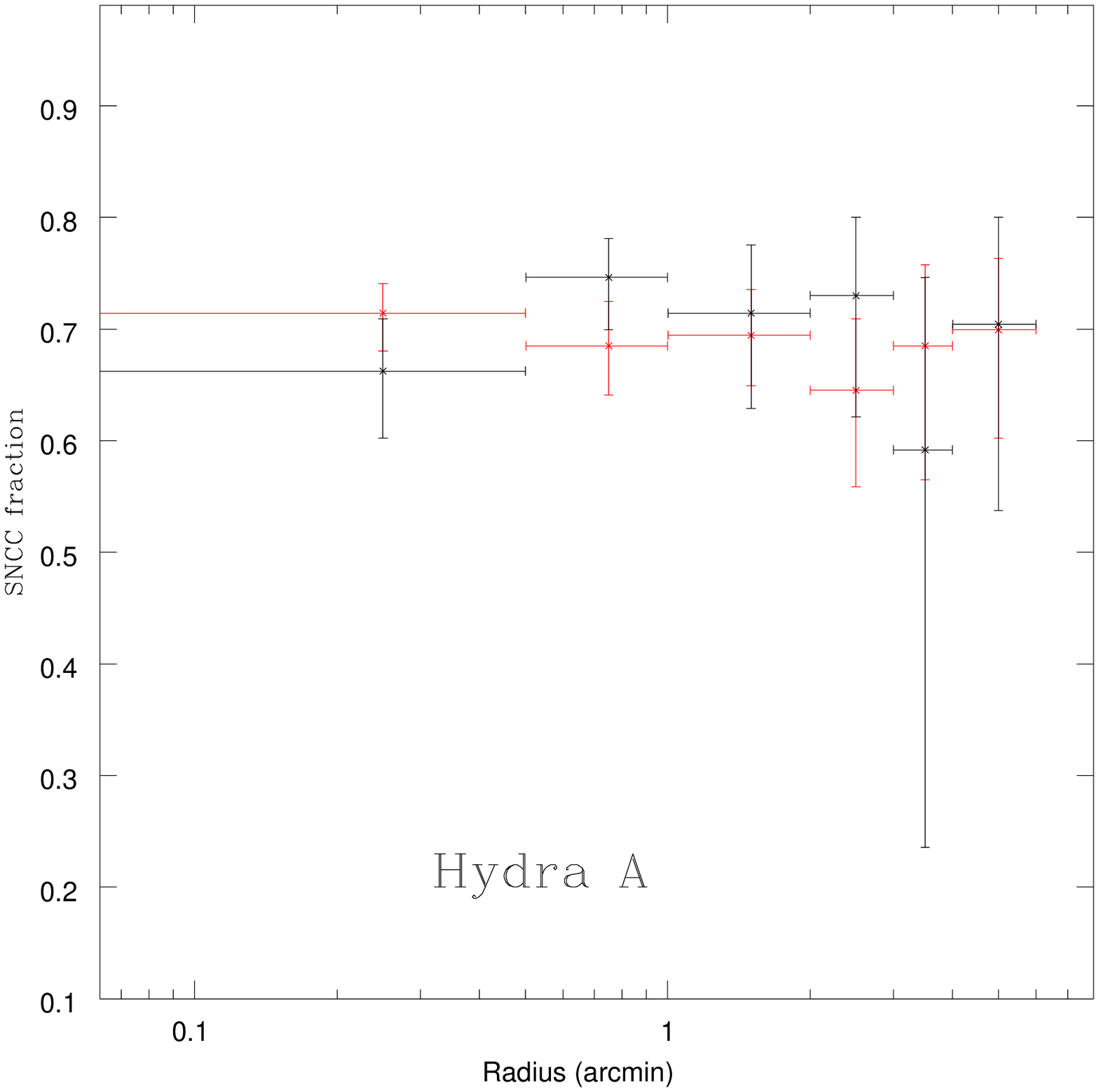,width=0.33\textwidth}  
      
}
\vspace*{-0.5ex}
\caption{Radial profiles of the relative fraction of SNCC using the
  O/Fe (red) and Si/Fe (black) ratio. }
\label{fig:snratio} 
\end{figure*}

\subsection{Number ratio of SNCC to SN Ia}
We investigated the SNe Ia and SNCC (including type Ib and Ic)
contribution to the ICM metals following the method presented in
\cite{2009A&A...508..191L}.  Using the SNe Ia of the WDD2 model (the
currently favoured SN Ia explosion scenario) adapted from
\citet{1999ApJS..125..439I} and SNCC yields by
\citet{1995MNRAS.277..945T} we computed which SNCC/Ia ratio reproduces
the observed abundances better. We note that although the models
adopted here (SNe yields, Salpeter IMF, spherical symmetry in the ICM,
etc.)  are probably too simplified and we should take into account
other effects (i.e metals locked in stars), we want to show if there
is agreement between the measured abundances and the theoretical
yields for different elements. Table \ref{table:snratio} and
Fig. \ref{fig:snratio} summarize the results. We found good agreement
in the O, Si and Fe abundances for A496, S\'ersic, and Centaurus for
which the relative fraction of SNCC seems to increase toward the
outskirts, and for Hydra A for which the relative fraction of SNCC
seems to be constant. Due to the large error bars in the SN
determination of A2029 it is difficult to say whether there is an
agreement between the same elements. The flatness profile of the
relative fraction of SNCC in Hydra A up to 370 kpc could be a
consequence of the mixing due to the central
AGN. \cite{2009A&A...493..409S} showed that the O/Fe in the cooler gas
extending in arm-like structures caused by the central AGN is
consistent with the average O/Fe ratio in the inner
3$^{\prime}$. Concerning A2029, we found an agreement between the O,
Si, and Fe only in a few regions probably due to the fact that the
spectral lines are weaker at high temperature and the determination of
the elements becomes difficult. We note that when the S/Fe ratio
increases the SNCC fraction computed using Si/Fe ratio is not in
agreement with what obtained using O/Fe ratio. Furthermore, even for
most of the considered regions in the other clusters, the S/Fe ratio
suggests a very low contribution to the enrichment due to SNCC that is
in contradiction with the idea of an early enrichment due to massive
stars. In other words, either the yields of sulfur from theoretical
works are underestimated or when the abundances of Si is high the
determination of sulfur becomes very difficult. We note that
\citep{2005ApJ...620..680B} found that clusters with a temperature
between 6 and 8 keV (as for A2029) show a general increasing of sulfur
corresponding to a decreasing of silicon.  \\ Apart from A2029, the
abundances of the other four clusters are consistent with a SNCC to SN
Ia ratio of around 1.5-3. This ratio means that almost 75-85$\%$ of
Fe, 15-25$\%$ of Si and $<20\%$ of S were synthetized by SN Ia.  Using
deep observations of 2A 0335+096 and S\'ersic 159-03
\cite{2006A&A...449..475W} and \cite{2006A&A...452..397D} found a
ratio of 2-2.5 in agreement with our result. Based on \xmm
observations of 22 clusters \cite{2007A&A...465..345D} found a ratio
in the range 1.7-3.5 depending on the supernovae models (using a WDD2
model, as in this work, they obtained 1.7 in good agrement with our
result). \\ The efficiency of the metal enrichment may depend on
parameters as age, star-formation efficiency and contribution from cD
galaxies. However, the relative contribution of SNe Ia and SNCC and
the processes of metal mixing in the ICM seem to be quite similar
among different clusters as claimed also by
\cite{2008arXiv0810.3820S}.  Our results are therefore consistent with
an enrichment due to the combination of SN Ia and SNCC. This
conclusion contrasts with the general views that the central iron
excess in cool core clusters is from Type Ia supernovae alone
(\citealt{2003A&A...401..443M}).

\section{2D Distribution of Metals}
\begin{figure*}[!hp]
\hbox{ 
      \epsfig{figure=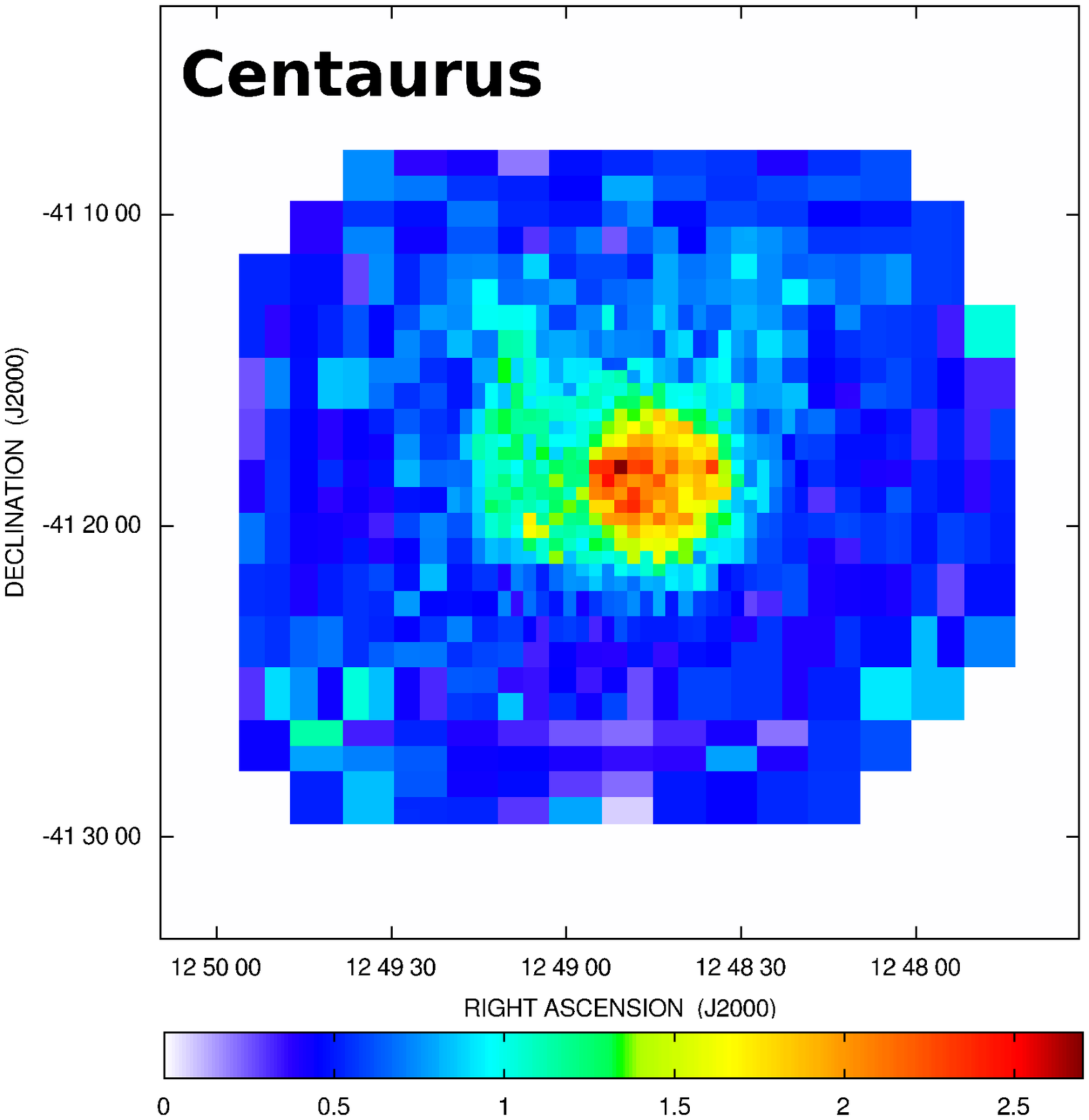,width=0.5\textwidth,height=0.33\textheight} 
      \epsfig{figure=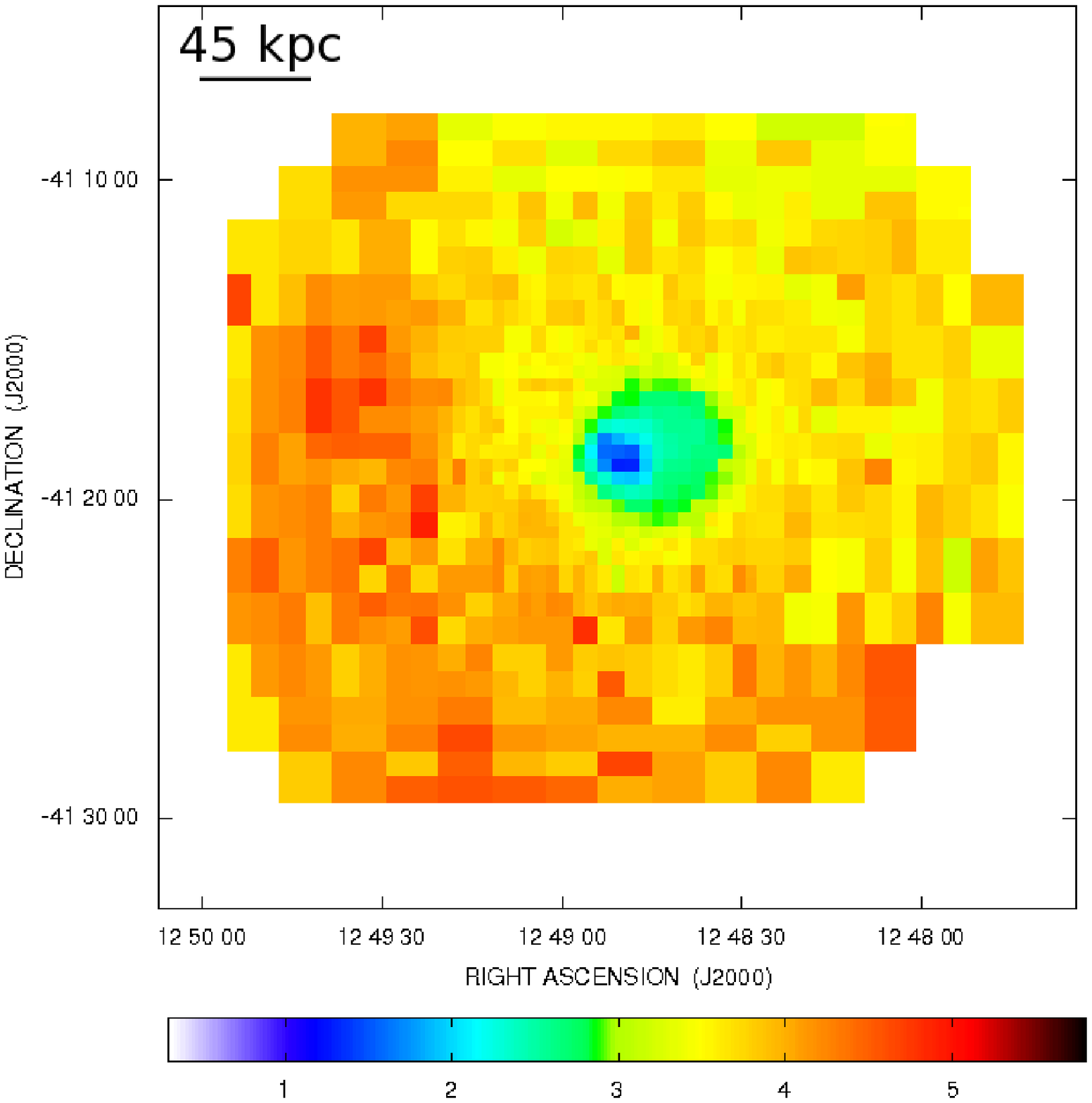,width=0.5\textwidth,height=0.33\textheight}  
}
\hbox{ 
      \epsfig{figure=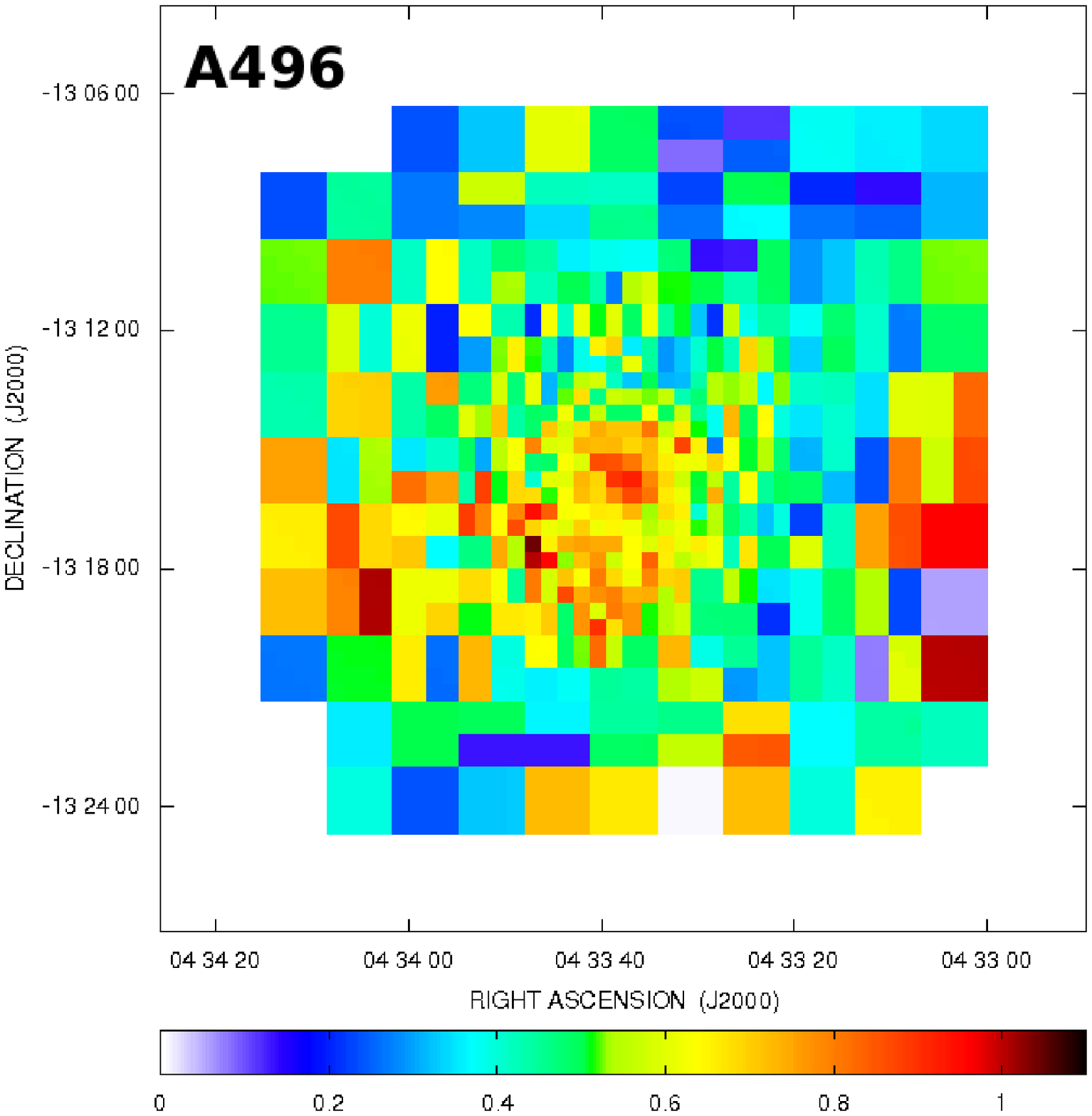,width=0.5\textwidth,height=0.33\textheight} 
      \epsfig{figure=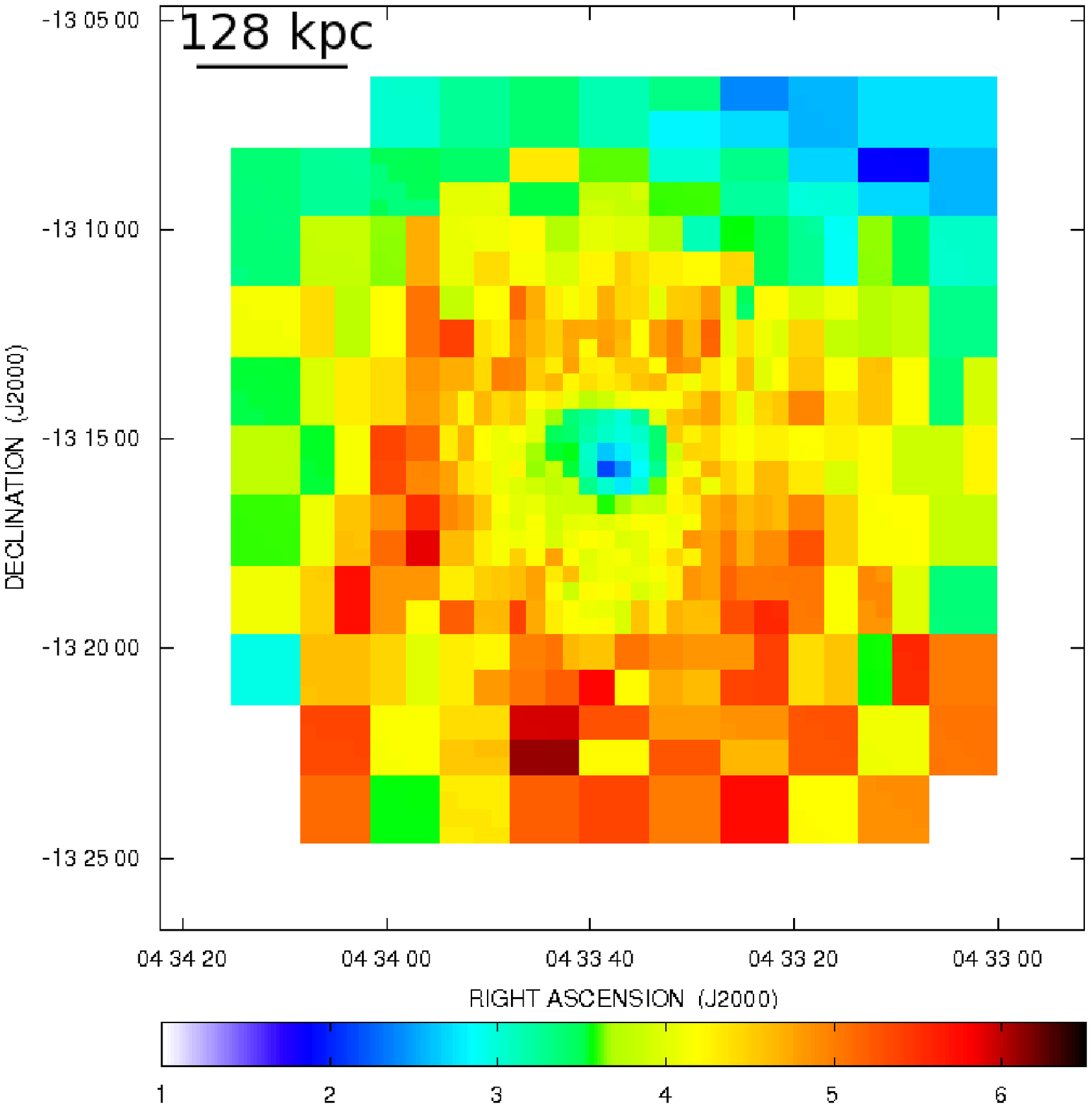,width=0.5\textwidth,height=0.33\textheight}  
}
\hbox{ 
      \epsfig{figure=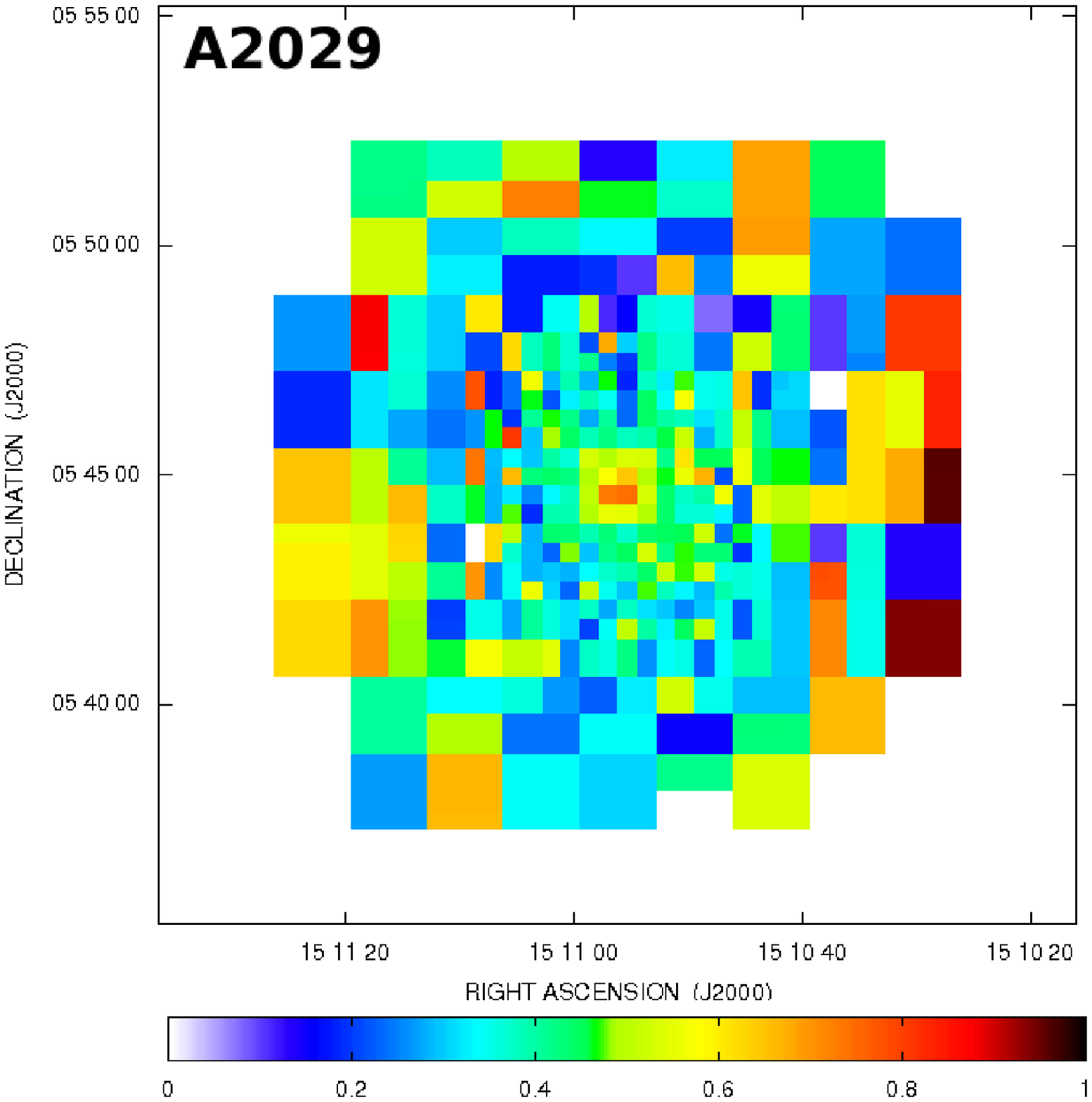,width=0.5\textwidth,height=0.33\textheight} 
      \epsfig{figure=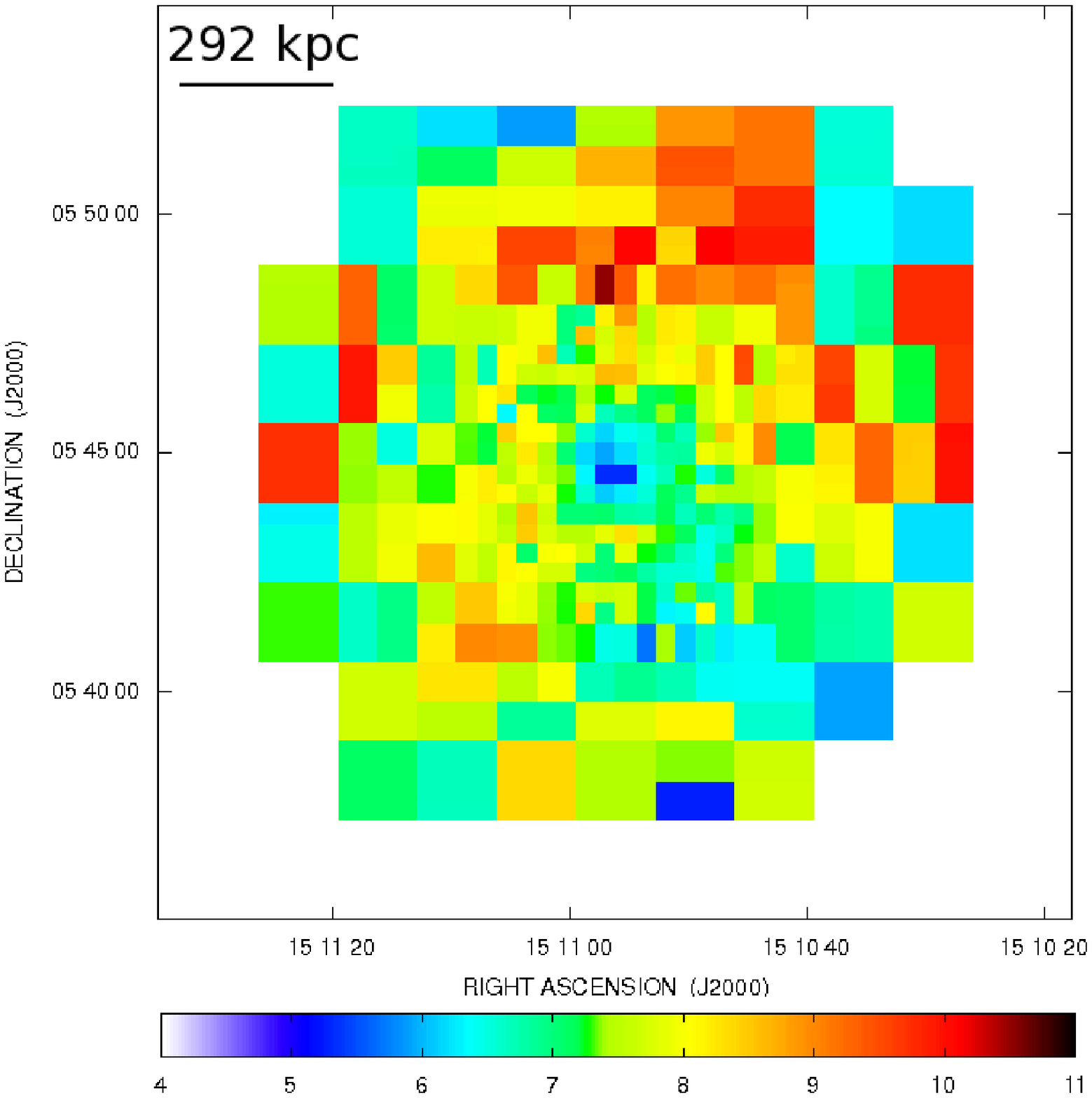,width=0.5\textwidth,height=0.33\textheight}  
}

\caption{{\it left:} Metallicity maps based on spectra from all three
  EPIC camera. {\it right:} Temperature maps, obtainded fitting the
  spectra with a single temperature model, with the same resolution of
  the metal maps. The scale for the metallicity and temperature is in
  solar units and keV, respectively. }
\label{fig:metalmaps} 
\end{figure*}

\begin{figure*}[t]
\hbox{ 
      \epsfig{figure=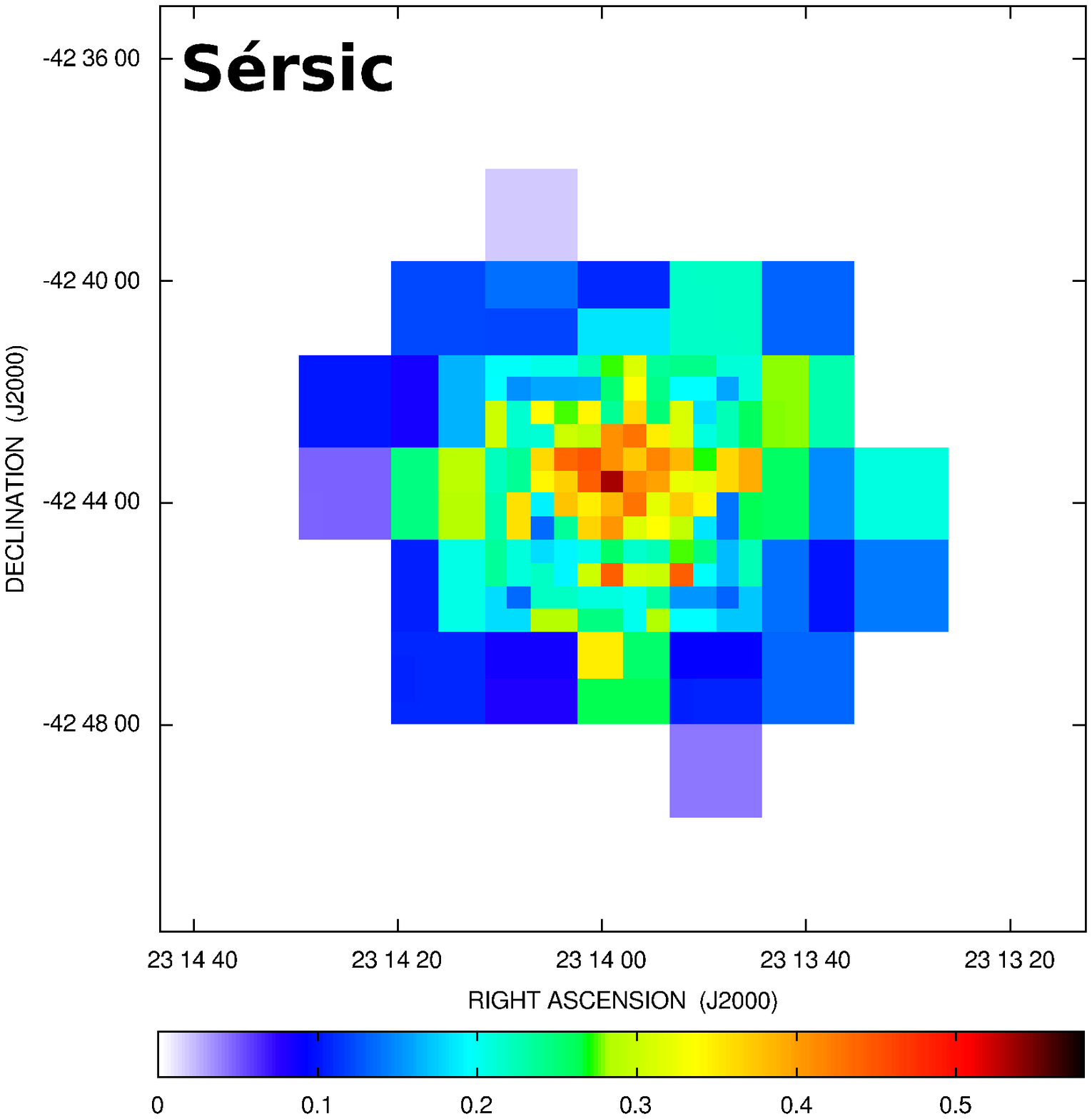,width=0.5\textwidth,height=0.33\textheight} 
      \epsfig{figure=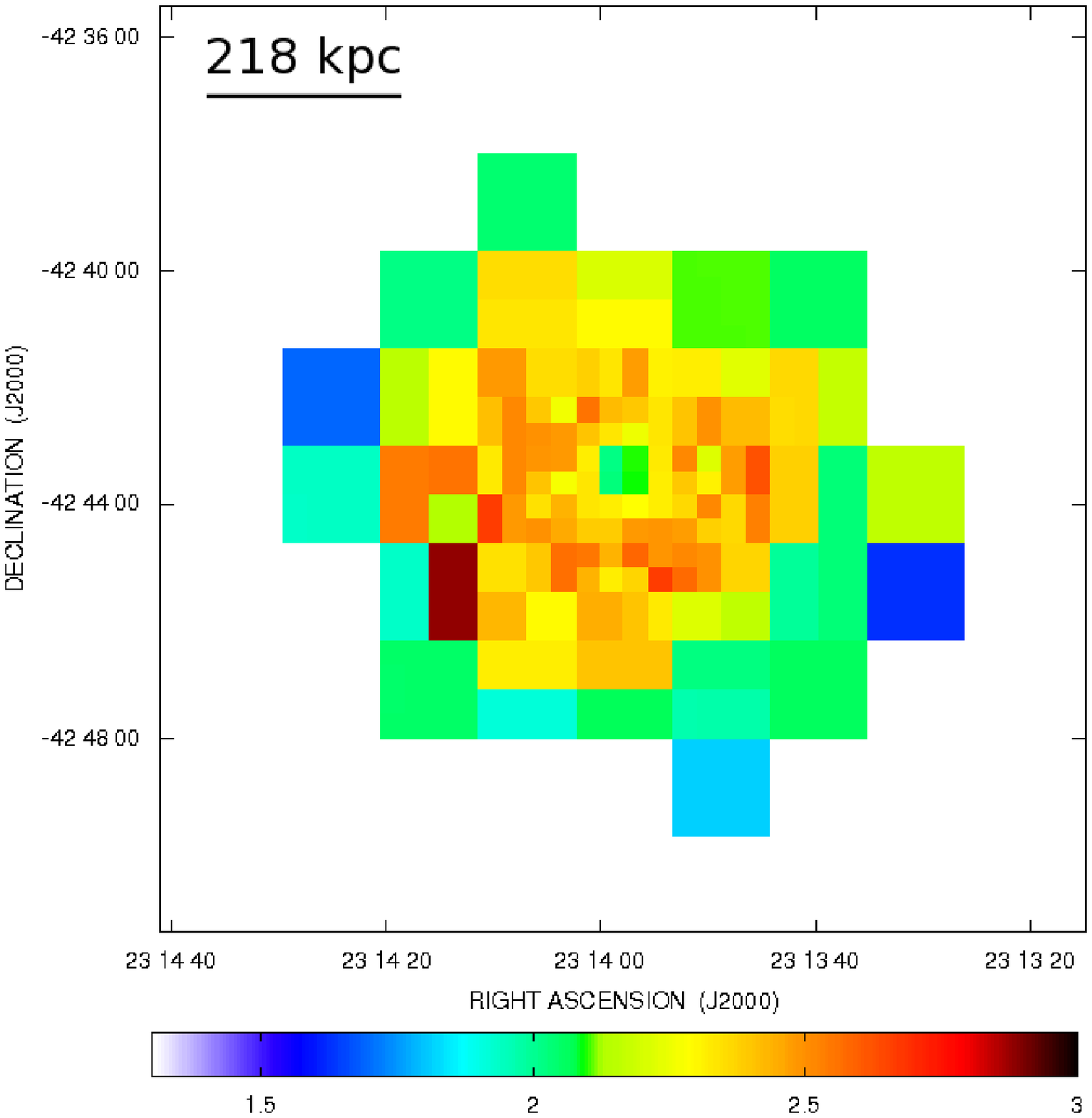,width=0.5\textwidth,height=0.33\textheight}  
}
\hbox{ 
      \epsfig{figure=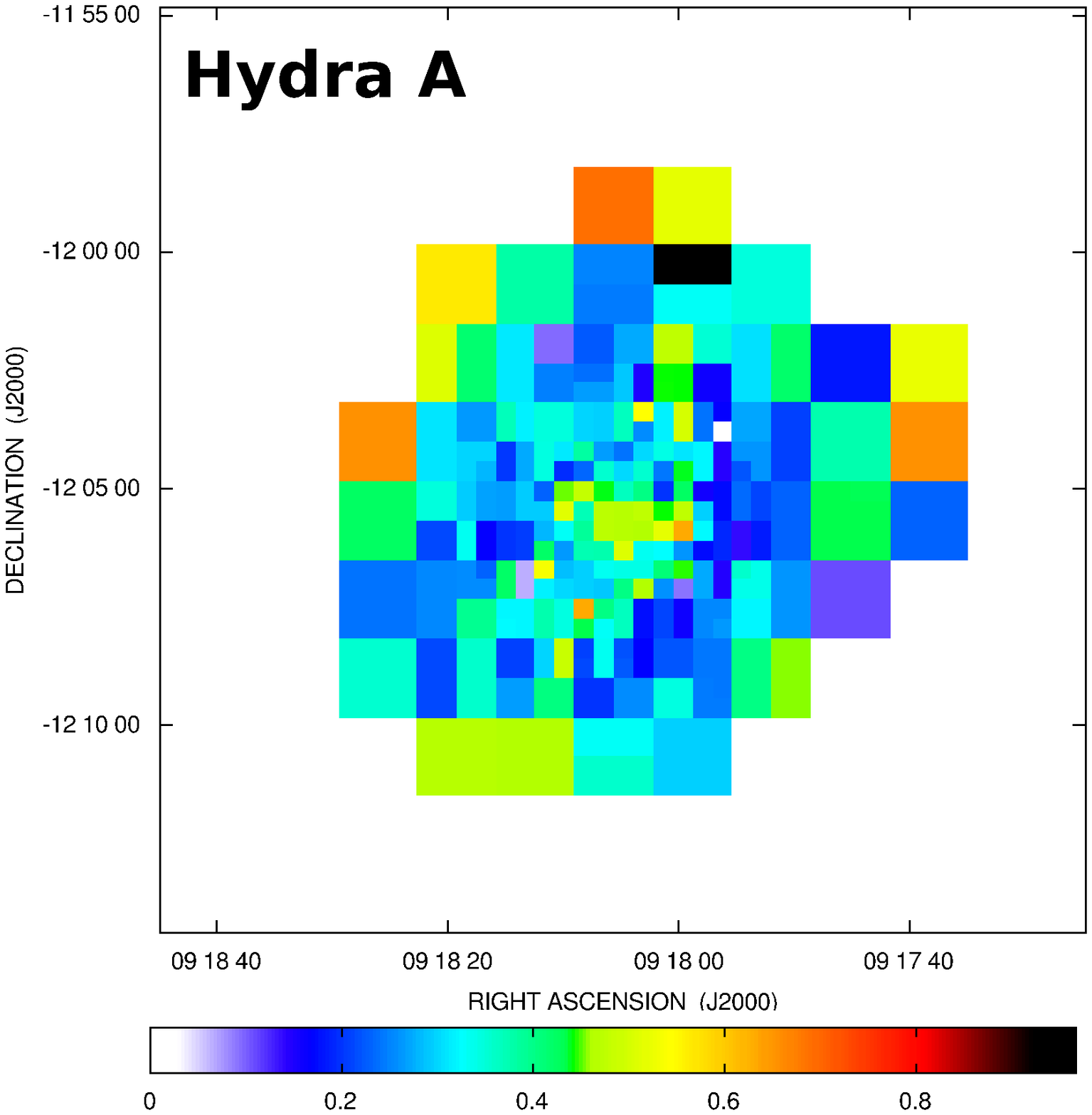,width=0.5\textwidth,height=0.33\textheight} 
      \epsfig{figure=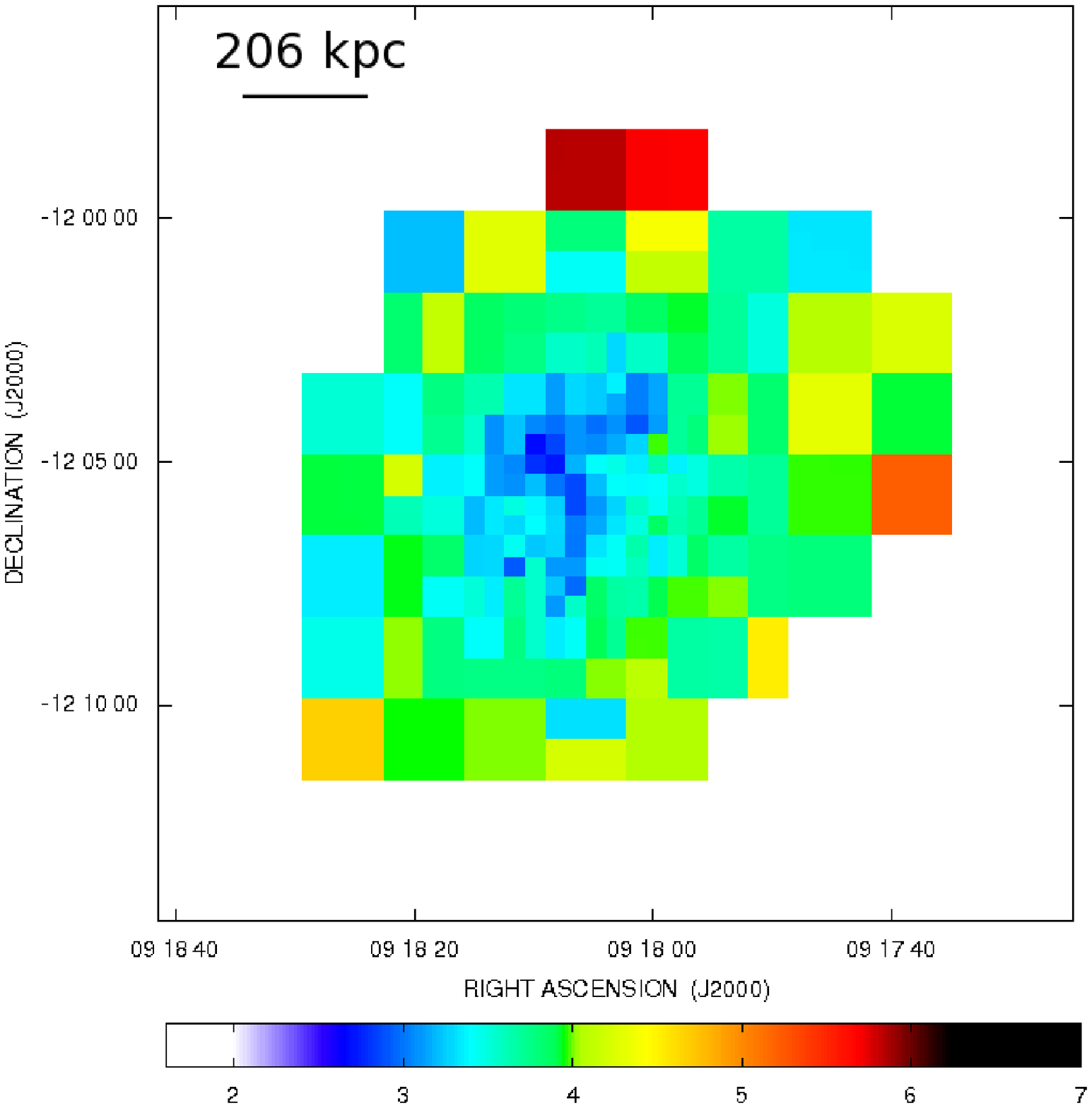,width=0.5\textwidth,height=0.33\textheight}  
}
\caption{Same as Fig. \ref{fig:metalmaps} } 
\label{fig:metalmaps2} 
\end{figure*}

\begin{figure*}[!ht]
\hbox{ 
      \epsfig{figure=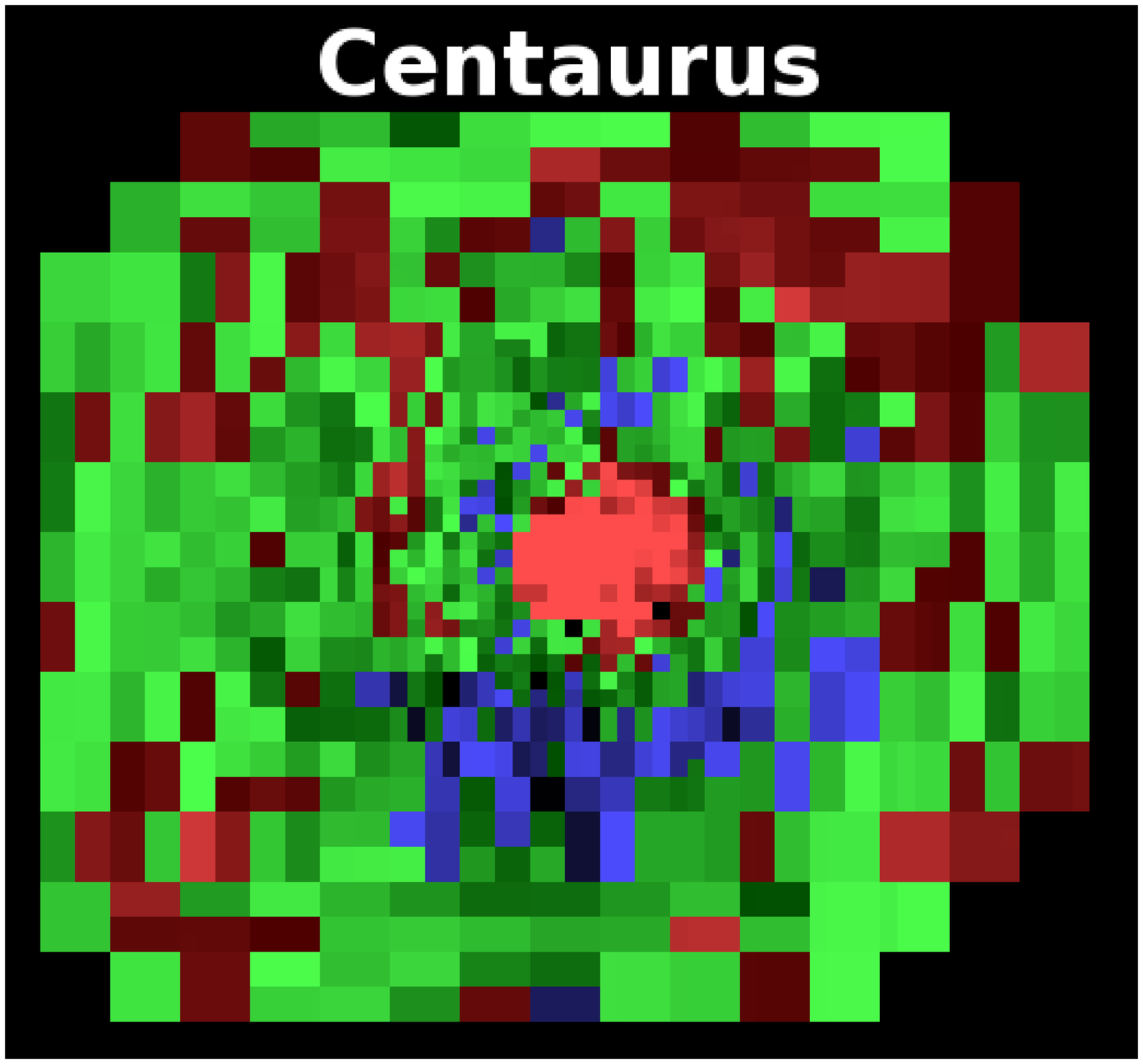,width=0.33\textwidth,height=5cm} 
      \epsfig{figure=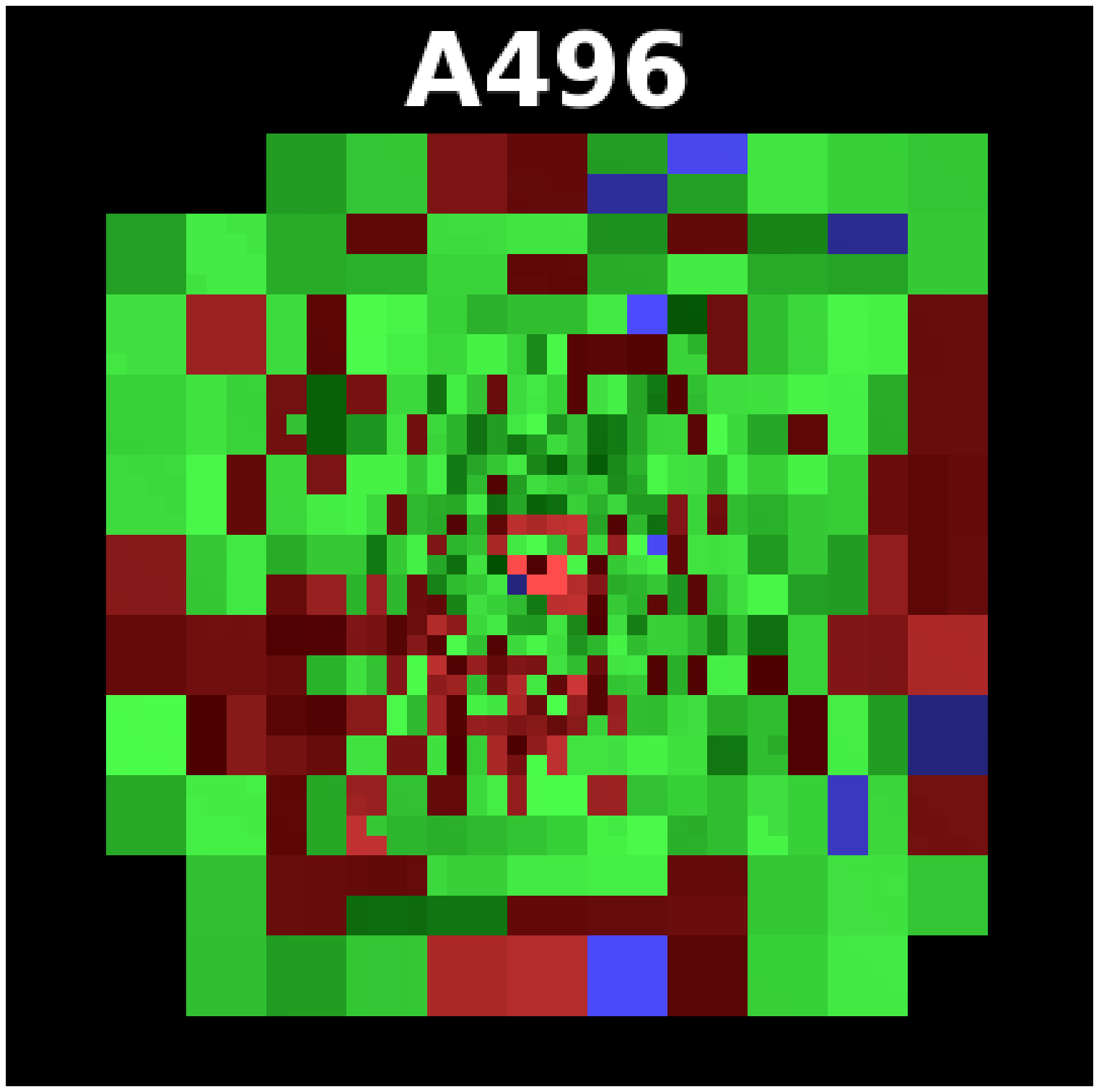,width=0.33\textwidth,height=5cm}  
      \epsfig{figure=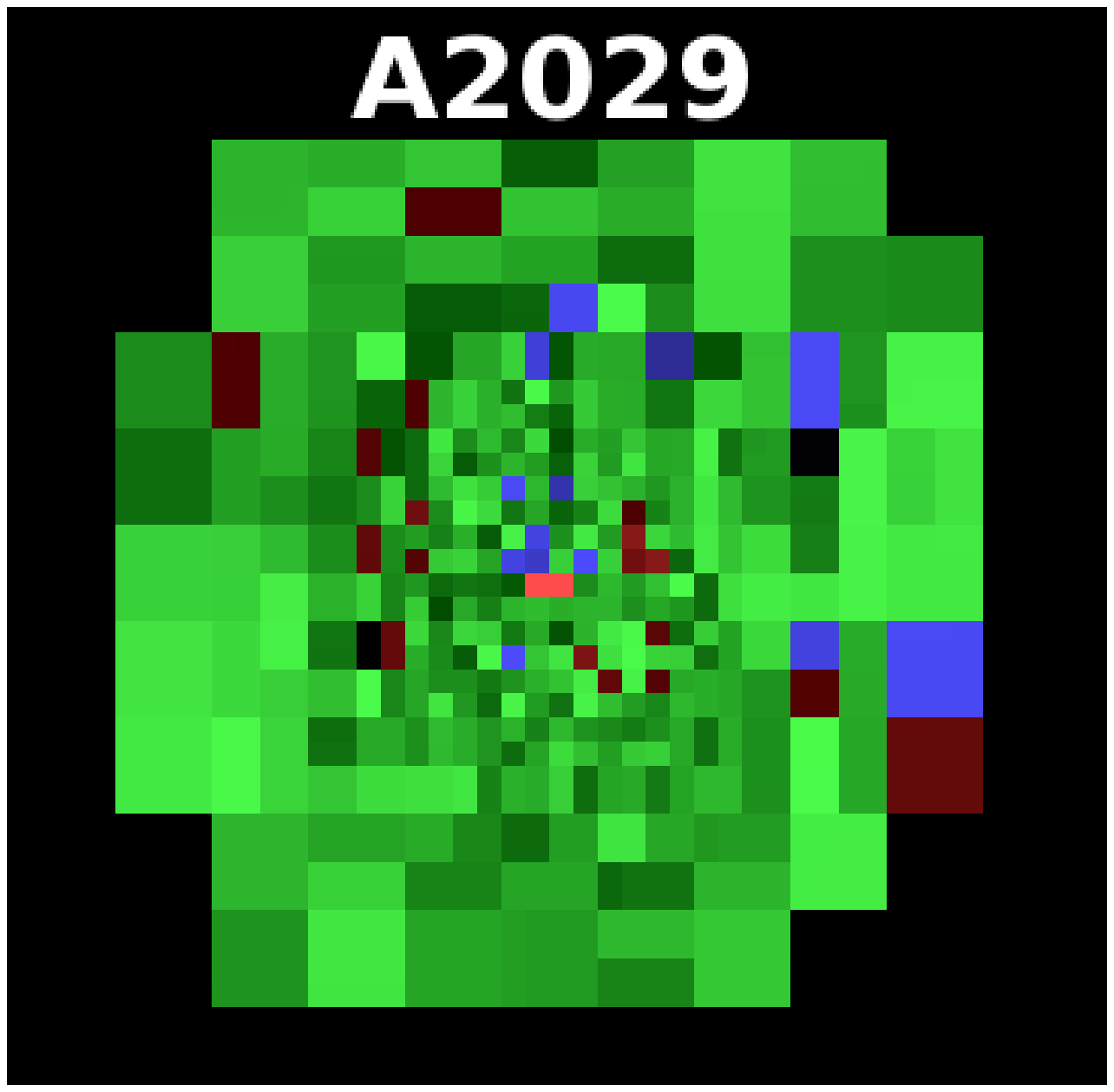,width=0.33\textwidth,height=5cm}  
}
\hbox{ 
      \epsfig{figure=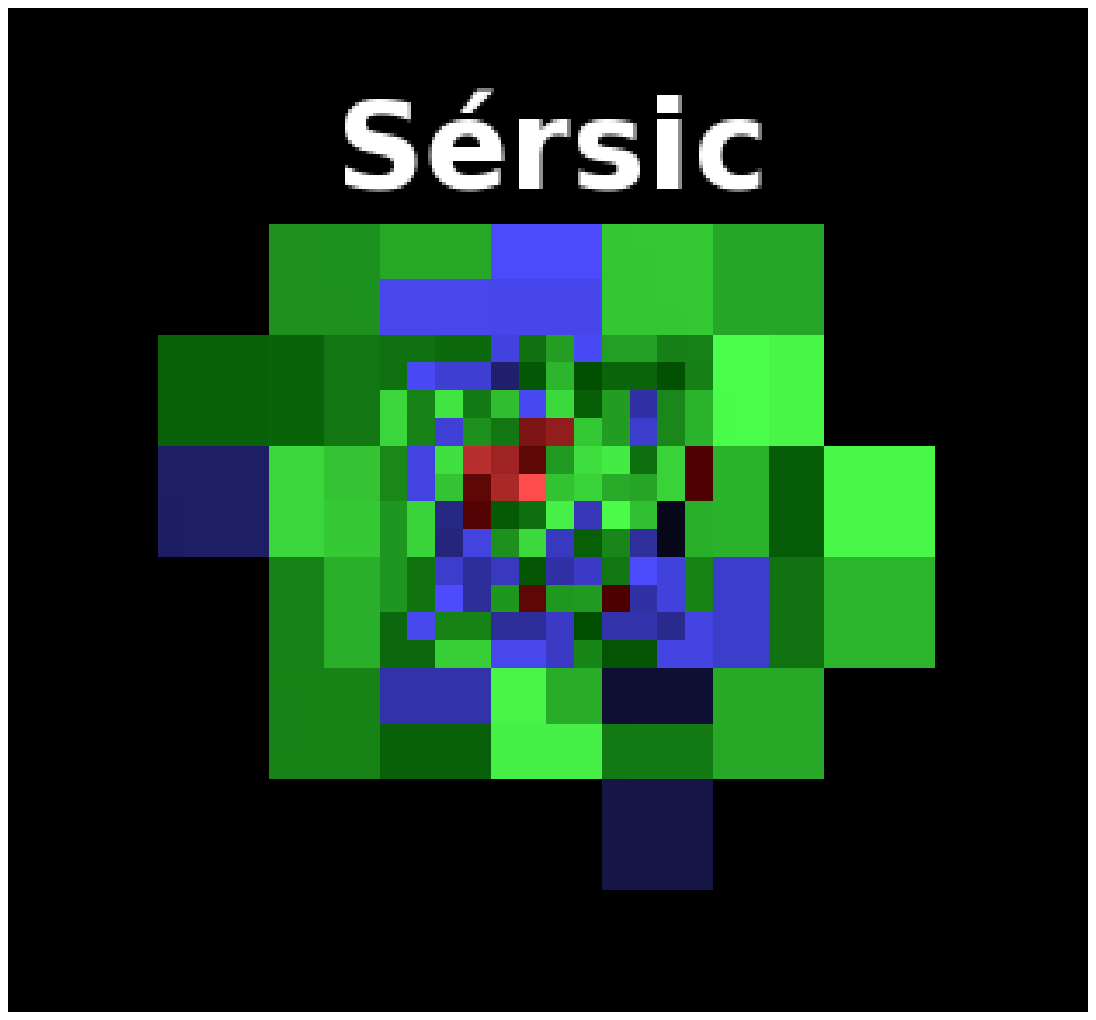,width=0.33\textwidth,height=5cm} 
      \epsfig{figure=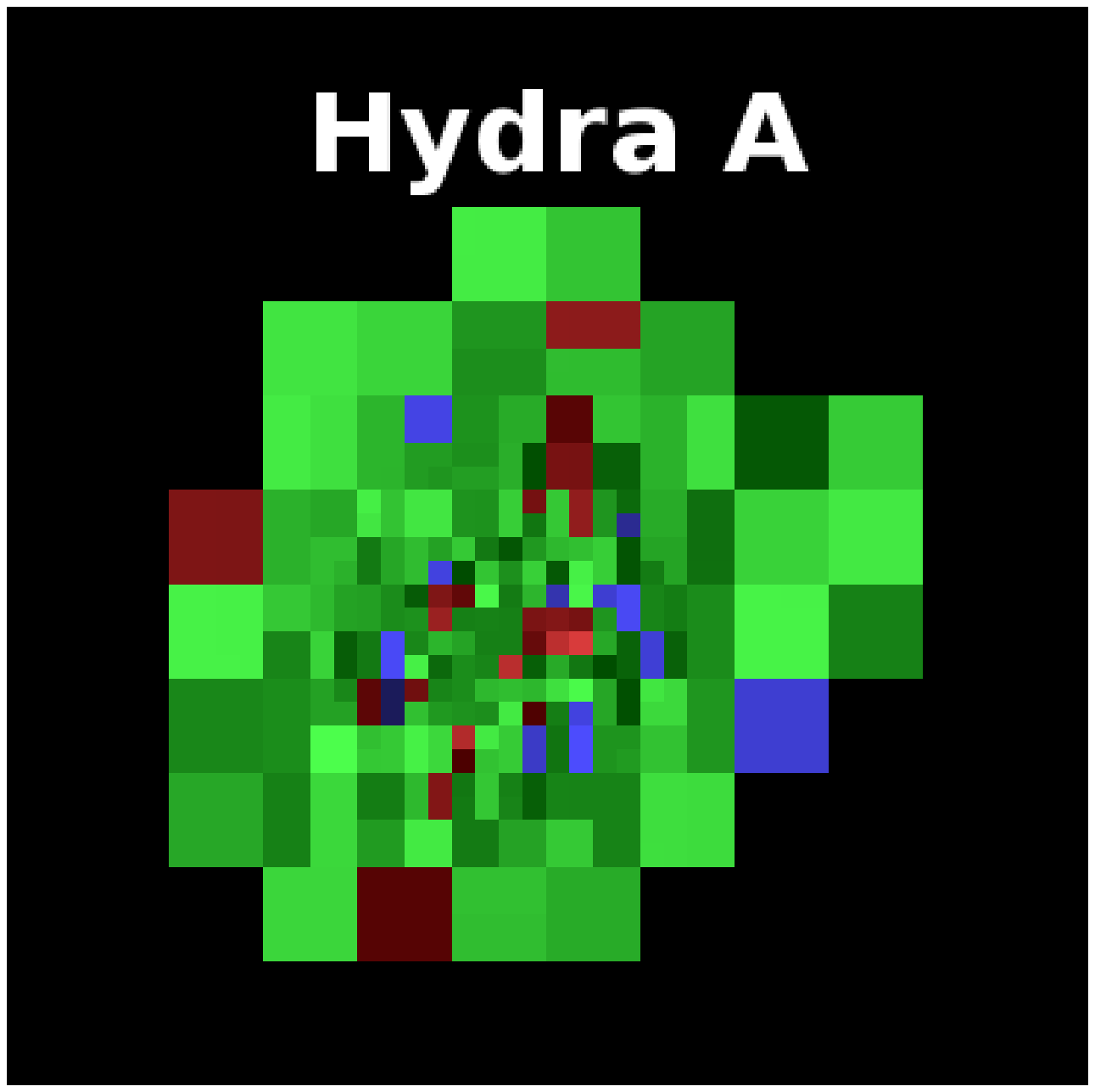,width=0.33\textwidth,height=5cm}
      \epsfig{figure=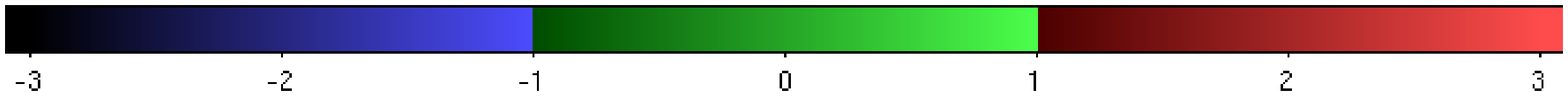,width=0.33\textwidth,height=0.035\textheight}     
}
\vspace*{-0.5ex}
\caption{Significance maps of all the clusters.}
\label{fig:significance} 
\end{figure*}

\begin{figure*}[!ht]
\hbox{ 
      \epsfig{figure=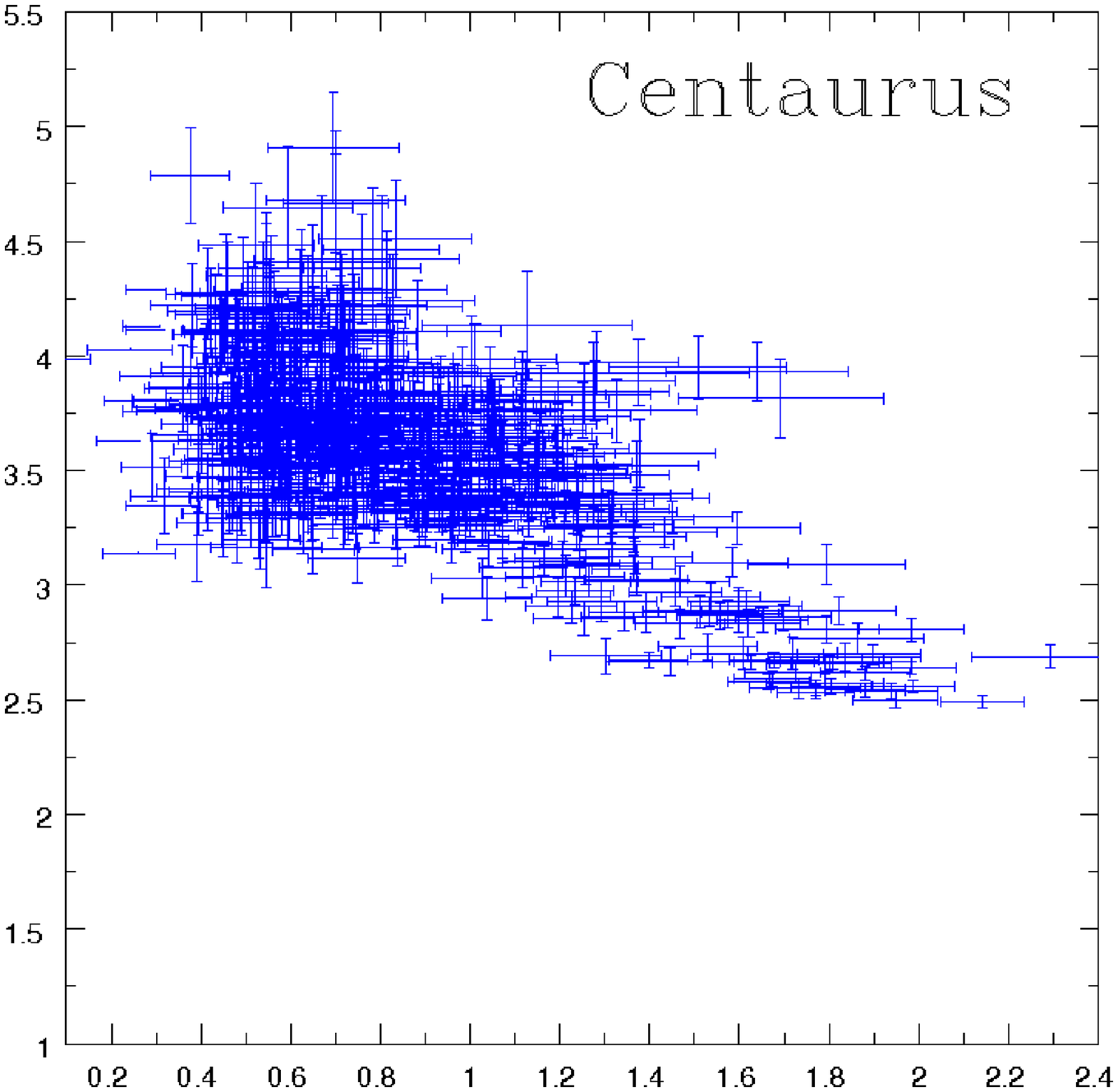,width=0.33\textwidth} 
      \epsfig{figure=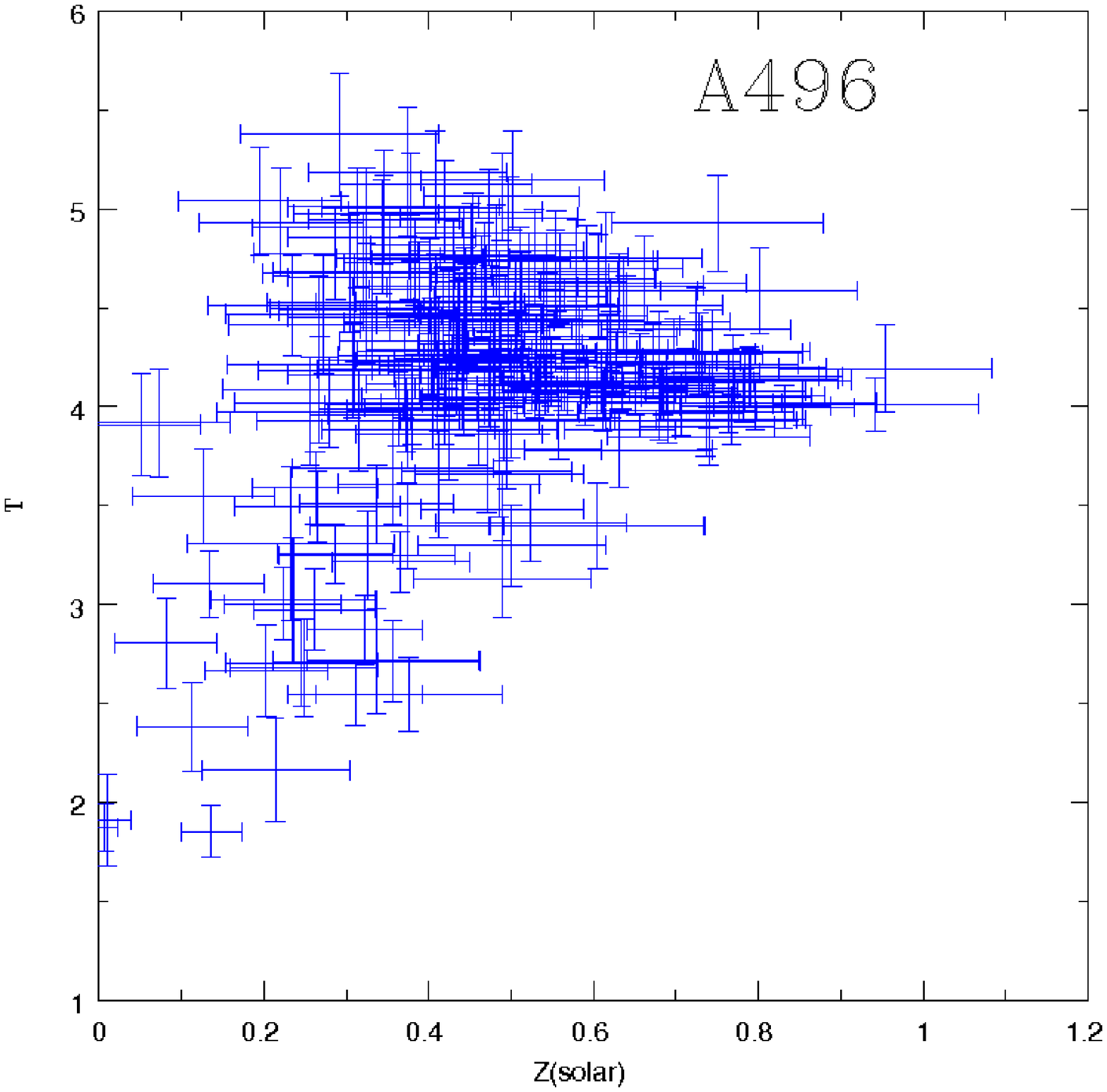,width=0.33\textwidth}  
      \epsfig{figure=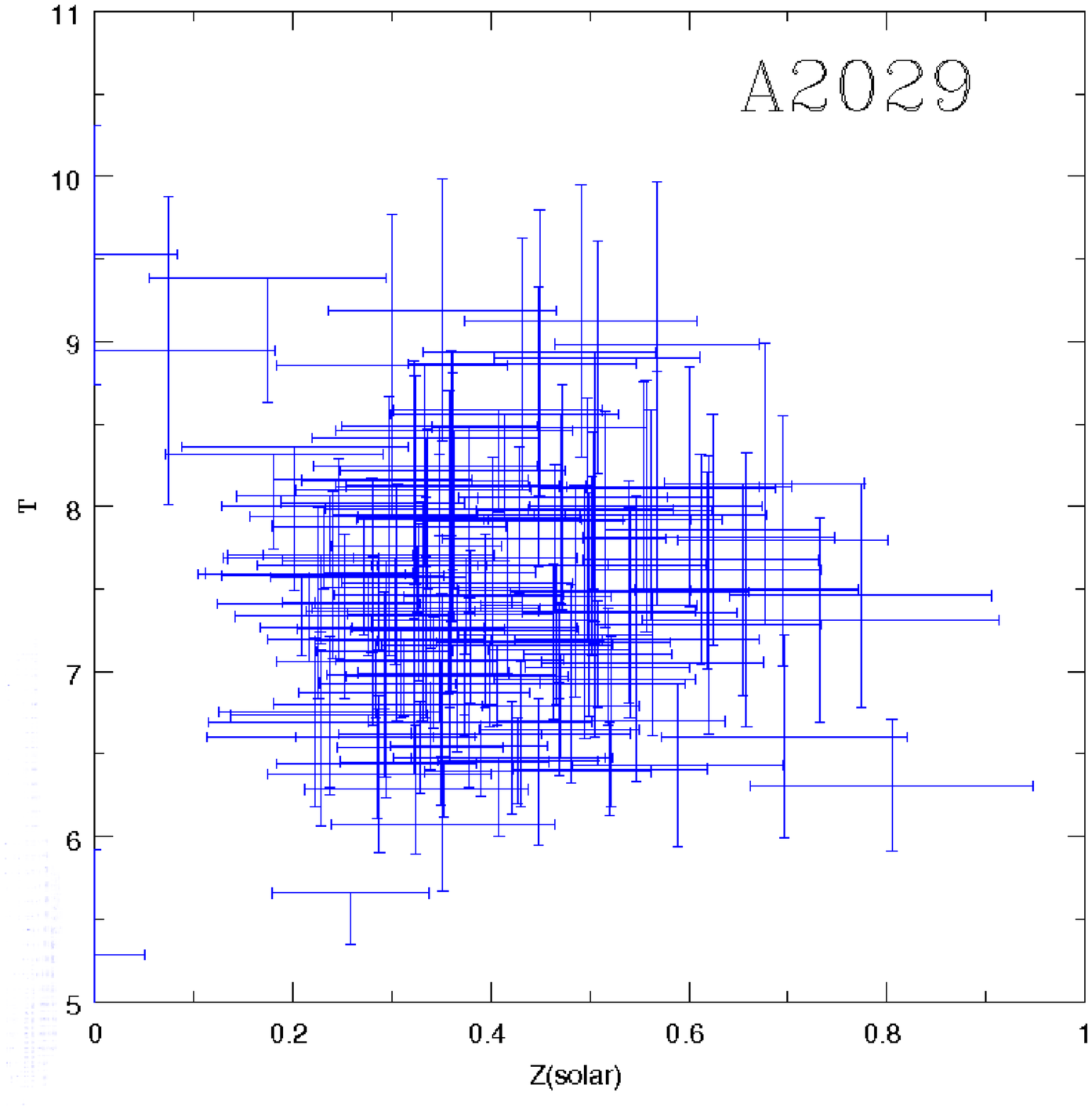,width=0.33\textwidth}  
}
\hbox{ 
      \epsfig{figure=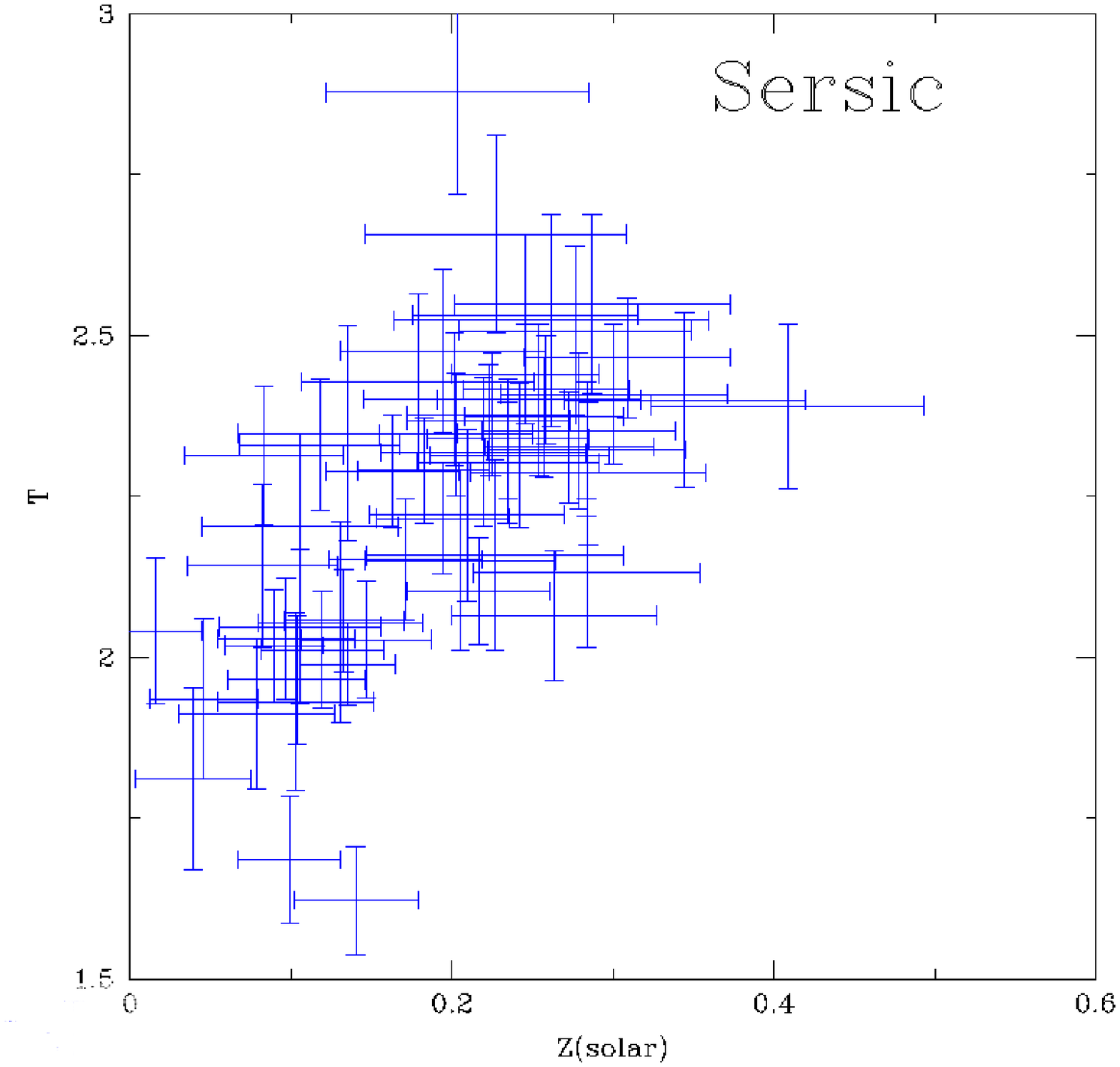,width=0.33\textwidth} 
      \epsfig{figure=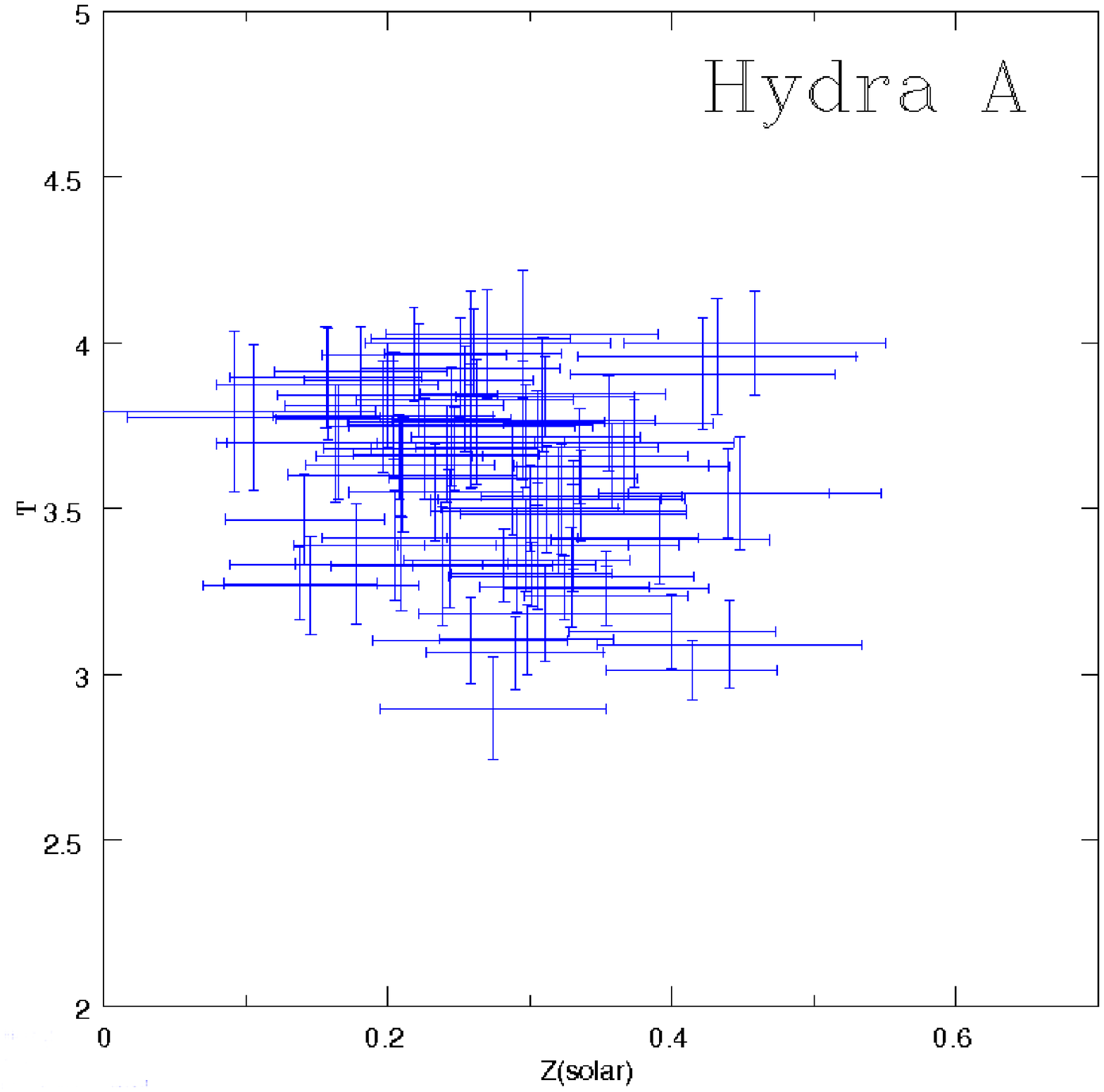,width=0.33\textwidth}
      }
\vspace*{-0.5ex}
\caption{Plot of abundance against temperature for each bin. }
\label{fig:ttzz} 
\end{figure*}

In order to study the distribution of metals, we prepared adaptively
binned abundance maps of the clusters. Appropriately cleaned data sets
for all the three cameras, with point sources removed were used to
create the source spectra. To obtain a metallicity measurement with a
good accuracy requires a high statistic. Thus, to ensure an acceptable
error also in the outskirts of the clusters we set a minimum count
number ($\sim$5000 source counts per region) necessary for proceeding
with the spectral fit. The spectral regions for the map were selected
following the method presented in \cite{2009A&A...508..191L} that we
can summarize in this way: a square region centered on the X-ray peak
was defined to include the area with high surface brightness. The
region size of the pixels was optimized to be as small as possible by
splitting it into horizontal or vertical segments through its center,
while including at least 5000 source counts. For all the selected
regions, spectra were extracted for source and background in all three
cameras. \\ The obtained metallicity maps are shown in
Fig. \ref{fig:metalmaps} and \ref{fig:metalmaps2}. The metallicity
distribution appears very inhomogeneous for all the clusters. For
Centaurus and S\'ersic 159-03, there is a peak in the center and then
it decrease in the outskirts while A496, A2029 and Hydra A show high
metallicity clumps both in the center and in the outskirts. We note
that, since Centaurus is at very low redshift, in these observations
we are looking in the very central part of the cluster (r$<$200 kpc)
compared with the others 4 clusters, and it could explain its
different shape. On the other hand, S\'ersic, for which we map the
metal distribution for more than 350 kpc we observe the same shape of
Centaurus. Since it is difficult to distinguish real metallicity
clumps from statistical noise we quantified the inhomogeneities
through the significance maps. First, we smoothed the metallicity
profiles applying a Savitzky-Golay filter and we subtracted it from
the metallicity maps. Finally we divided each bin of the resultant
maps by the uncertainty in the pixel metallicity. The results are
shown in Fig \ref{fig:significance}. The blue and red spots represent
regions that deviate significatively (99$\%$ c.l.) from the average
profile. \\
\noindent There are several maxima visible in the metal distribution,
which are not associated with the cluster center. From simulations
(\citealt{2008MNRAS.389.1405K, 2009A&A...499...87K}) we know that the
maxima are typically at places where galaxies just have lost a lot of
gas due to the ram-pressure. Since the gas lost by galaxies is
obviously not mixed immediately with the ICM at the place where we
observe a metal blob we should observe also a low temperature due to
the fact that the gas in galaxies is cooler than the ICM. Thus, we
produced the temperature maps (fitting the spectra with a single
temperature model) of the clusters with the same spatial resolution
obtained for the metal maps and then we plotted the abundance of bins
against their temperature, which are shown in
Fig. \ref{fig:ttzz}. Since we are searching for cool high-metallicity
clumps due to the ejection of gas from galaxies and not the cool
high-metallicity bins found in the cool cores, we did not plot the
inner bins (where we used a two temperature model to fit the
spectra). \\ Apart from Centaurus, we see a deviation from the
expected temperature-metallicity relation. The deviation could be due
to the combination of the iron bias and inverse iron bias effects. The
iron bias effects cause an underestimation of the metallicity when we
pretend to fit with a single temperature model a plasma that is,
instead characterized by a combination of different temperatures
(\citealt{2000MNRAS.311..176B}, \citealt{2008ApJ...674..728R}). On the
contrary the inverse iron bias effect cause an overestimation of the
metal abundance (\citealt{2010arXiv1006.3255G}). The combination of
these two effects could explain the spread in the distribution of
Fig. \ref{fig:ttzz}. On the other hand these effects seems to be more
important in the range of temperature between 2 and 4 keV, while our
clusters show also regions with temperature higher than 4 keV.  One
possible explanation could be that the ejected gas, with $T<1$ keV and
metallicity in the range 0.5-1.5 (\citealt{2000PASJ...52..685M},
\citealt{2009ApJ...696..681A}), will be heated up to the temperature
of the surrounding gas (ICM) on the shorter time-scale than that of
metal mixing. In this case, after a while we should observe a region
of high metallicity (not yet dispersed) and high temperature (heated
up at the ICM temperature). Another possible explanation could be
related to the number of intracluster supernovae. In fact during
ram-pressure stripping events a lot of stars are forming in the tail
of stripped gas. The stars are evolving and exploding as SNe directly
in the ICM and they can enrich the ICM very efficiently. In this case
we should see a clumps of high metallicity (due to SNe explosions) and
high temperature.  Obviously, more complex heating and cooling
processes are at work, thus the simple picture of stripped gas does
not hold.

\section{Mass determination}
Using simulated galaxy clusters \cite{2007A&A...472..757K} showed that
the more inhomogeneous the metals are distributed within the cluster,
the more underestimated is metal mass. They showed that the true metal
mass in the inner parts (r$<$500 kpc) of galaxy cluster can be up to
three times higher than the metal mass obtained by X-ray
observations. They suggested that the discrepancies are due to the
fact that the metallicity is not constant throughout the extraction
area, thus the integration of thermal bremsstrahlung and of line
emission can lead to underestimated metal masses. \\
\noindent We used the
metallicity maps to estimate the metal mass in the center parts of the
clusters and compare the results with the estimations obtained from a
single extraction area of the cluster. \\
First, we computed a background subtracted, vignetting corrected,
radial surface brightness profile in the 0.3-10 keV energy band for
each cluster. All of the X-ray point sources were escluded from the
data. The annuli were chosen such that all the widths are larger than
the FWHM of the point-spread function (PSF) at that radius. With this
choice, all the bins contain at least 2000 counts after background
subtraction. With this very good statistics, the error bars of the
surface brightness are very small. The profiles were
fitted using a $\beta$-model \citep{1976A&A....49..137C}:
\begin{equation}
S(r)=S_0 \left( 1+\frac{r^2}{r_c^2} \right)^{-3\beta+0.5}
\end{equation}
where $r_c$ is the core radius. The advantage of using a $\beta$-model
to parametrize surface brightness is that assuming hydrostatic
equilibrium and spherical symmetry the gas density and total mass profile can be
recovered analytically and expressed by the simple formula:
\begin{equation}
n_{gas}(r)=n_0(1+x^2)^{-3\beta/2}
\end{equation}

\begin{equation}
M_{tot}(<r)=\frac{3\beta T_{gas} r_c}{G\mu m_p}\frac{x^3}{1+x^2}
\end{equation}
where $x=r/r_c$, $n_0$ is the central electron density, $\mu$ is the
mean molecular weight in atomic mass (=0.6), $G$ is the gravitational
constant and $m_p$ is the proton mass. The best-fit parameters
obtained from the spectral and spatial analysis are shown in Table
4. We then evaluate R$_{\Delta}$ as the radius
encompassing a fixed density contrast with respect to the critical
density $\rho_c$. This is necessary to compare different clusters.
Since Centaurus, is at very low redshift with the analyzed \xmm
observation we are looking at an overdensity of 4500. Thus we used
such overdensity to estimate the mass of the clusters. For the other
four clusters we estimated also the parameter at an overdensity of
2500 and at an overdensity corresponding to the area covered by the
metal maps.  We compute than the metal mass as :
\begin{equation} \label{eq:metals}
M_{metals}=M_{gas}Zf_{metals,\odot}
\end{equation}
where $Z$ is the metallicity of the gas and $f_{metals,\odot}$ is the
metal mass fraction of the Sun. The results are shown in Table 4.

\setlength{\extrarowheight}{0.2cm}
\begin{table*} \label{table:mass}
\caption{Best-fit results of the spectral and spatial analysis of the
  sample of galaxy clusters at an overdensity corresponding to
  4500. We determined the total mass using the T derived fitting the
  spectra with a single temperature model. The last column represents
  the ratio between column 10 and 9 containing the metal mass. }

            \begin{tabular}{c c c c c c c c c c c}
            \hline
            \noalign{\smallskip}
%   &    & &    &   &    &  & \multicolumn{2}{c}{$\beta$-model}  &  \multicolumn{2}{c}{metal \ maps} & & \\
        &          &              &       &          &          &                  &                    & $\beta$-model     & metal maps         &  \\
Cluster & kT$_{gas}$ & Z/Z$_{\odot}$ & r$_c$ & $\beta$  & $\Delta$ & M$_{tot}$         & M$_{gas}$          & M$_{metals}$       & M$_{metals}$        & ratio  \\
        &   keV    &              &  kpc  &          &          & 10$^{13}M_{\odot}$ & 10$^{12}M_{\odot}$  &  10$^{10}M_{\odot}$ &  10$^{10}M_{\odot}$ & C. 10 to C. 9 \\
            \noalign{\smallskip}
            \hline
            \noalign{\smallskip}
            S\'ersic   & 2.38$\pm$0.01 & 0.357$\pm$0.003 & 37$\pm$1 & 0.602$\pm$0.001 & 4500 & 2.59$\pm$0.18  & 2.45$\pm$0.12 & 1.75$\pm$0.09 & 1.92$\pm$0.25 & 1.34$\pm$0.23 \\    
            Centaurus           & 3.74$\pm$0.01 & 0.810$\pm$0.003 & 17$\pm$1 & 0.397$\pm$0.001  & 4500 & 2.91$\pm$0.14  & 1.20$\pm$0.11 & 1.94$\pm$0.19 & 2.17$\pm$0.18 & 1.12$\pm$0.19 \\
            Hydra              & 3.57$\pm$0.01 & 0.357$\pm$0.005 & 35$\pm$1 & 0.561$\pm$0.001  & 4500 & 3.76$\pm$0.22  & 3.36$\pm$0.20 & 2.40$\pm$0.14 & 3.32$\pm$0.59 & 1.38$\pm$0.31 \\
 A496                & 3.72$\pm$0.01 & 0.586$\pm$0.004 & 27$\pm$1 & 0.459$\pm$0.001  & 4500 & 3.76$\pm$0.19  & 2.72$\pm$0.16 & 3.19$\pm$0.18 & 3.91$\pm$0.45 & 1.23$\pm$0.20 \\
            A2029               & 7.01$\pm$0.02 & 0.493$\pm$0.006 & 41$\pm$1 & 0.563$\pm$0.001  & 4500 & 12.26$\pm$0.81  & 11.42$\pm$0.83 & 11.79$\pm$1.01 & 13.97$\pm$0.95 & 1.18$\pm$0.16 \\
            \hline
            S\'ersic   & 2.39$\pm$0.01 & 0.309$\pm$0.003 & 37$\pm$1 & 0.602$\pm$0.001 & 2500 & 3.95$\pm$0.11 & 3.95$\pm$0.21 & 2.44$\pm$0.13 & 3.03$\pm$0.21 & 1.24$\pm0.19$ \\    
            Hydra               & 3.46$\pm$0.01 & 0.346$\pm$0.005 & 35$\pm$1 & 0.561$\pm$0.001 & 2500 & 5.47$\pm$0.29 & 5.53$\pm$0.41 & 3.83$\pm$0.43 & 4.97$\pm$0.55 & 1.30$\pm$0.26 \\
            A496                & 3.80$\pm$0.01 & 0.506$\pm$0.004 & 27$\pm$1 & 0.459$\pm$0.001 & 2500 & 5.30$\pm$0.20 & 4.79$\pm$0.31 & 4.85$\pm$0.31 & 6.38$\pm$0.70 & 1.32$\pm$0.22 \\
            A2029               & 7.07$\pm$0.02 & 0.468$\pm$0.006 & 41$\pm$1 & 0.563$\pm$0.001 & 2500 & 17.5$\pm$0.90 & 21.8$\pm$1.30 & 20.4$\pm$1.20 & 23.7$\pm$2.80 & 1.16$\pm$0.19 \\
            \hline
            A496                & 4.43$\pm$0.01 & 0.486$\pm$0.004 & 27$\pm$1 & 0.459$\pm$0.001 & 1600 & 7.78$\pm$0.36  & 6.79$\pm$0.33 & 6.60$\pm$0.38 & 8.79$\pm$1.00 & 1.33$\pm$0.21 \\
            Hydra               & 3.64$\pm$0.01 & 0.299$\pm$0.005 & 35$\pm$1 & 0.561$\pm$0.001 & 1400 & 8.35$\pm$0.44  & 10.29$\pm$0.56 & 6.15$\pm$0.34 & 7.63$\pm$0.99 & 1.24$\pm$0.22 \\
            S\'ersic            & 2.31$\pm$0.01 & 0.286$\pm$0.003 & 37$\pm$1 & 0.602$\pm$0.001 & 1200 & 5.05$\pm$0.34  & 6.79$\pm$0.29 & 3.88$\pm$0.29 & 4.83$\pm$0.32 & 1.24$\pm$0.16 \\    
            A2029               & 7.81$\pm$0.02 & 0.420$\pm$0.005 & 41$\pm$1 & 0.563$\pm$0.001 & 1200 & 34.0$\pm$1.80  & 25.3$\pm$1.30 & 21.3$\pm$1.04 & 49.3$\pm$9.90 & 2.31$\pm$0.55 \\
            \hline

                        \noalign{\smallskip}
            \hline
            \end{tabular}         
\end{table*}

\subsection{Metal mass estimation using metallicity maps}
In the ionized intra-cluster
plasma the ratio between the proton density n$_p$ and the electron
density n$_e$ is approximately 0.82. Thus, the emission integral (EI)
could be written as:
\begin{equation}
EI=\int n_pn_edV \approx n_pn_eV \approx 1.2 n_pV.
\end{equation}
For each spectrum the emission integral (EI) can
be derived easily using the normalization K of the thermal spectrum
measured within XSPEC:
\begin{equation}
EI=K\times10^{14}[4\pi d^{2}_{ang}(1+z)^2].
\end{equation}
For each pixel the gas mass along the line of sight is determined using:
\begin{equation}
M_{gas}=\sum_i m_in_iV \approx (m_Hn_H + m_{He}n_{He})V\approx 1.3m_H\sqrt{EI}\sqrt{V}
\end{equation}
where $n_H$ and $n_{He}$ are the proton and helium number density
respectively, $m_H$ is the proton mass, $m_{He}$=4$m_p$ and V is the
volume of the emitting region. This was determined (assuming that
  the properties of the material in each region are constant and that
  there is no material projected onto them) as $V\approx 2\sqrt{R^2-X^2-Y^2}A$,
  where $A$ is the area of the region, $R$ is the radius emcompassing
  the fixed density contrast, and $X$ and $Y$ are the projected
  distances in the east-west and north-south directions,
  respectively. We assumed the solar $He/H$ fraction,
$n_{He}/n_{H}\approx 0.095$ and we did not consider the mass
contribution of ions heavier than $He$ which are negligible compared
to $H$ and $He$. Then using the equation \ref{eq:metals} we determined
the metal mass along the line of sight for each pixel and summed them up
for all the pixels.\\
\noindent We found that the metal mass is higher than the metal mass
obtained by assuming the metallicity mean of the investigated area
(see Table 4).  The explanation is that when we
determine the metal mass using the mean metallicity we do not take
into account properly that there are a lot of metals at positions of
low density.  Since, the metallicity is not constant throughout the
galaxy cluster, as shown with the metallicity maps, using the maps we
are estimating the metal mass along the line of sight better. The
discrepancy changes from cluster to cluster, and considering different
radii. In the very inner part, at an overdensity of 4500 the metal
mass can be underestimated up to 30$\%$. At larger radii the
discrepancies can be up to more than 2
times. \\ 
From simulations we know that ram-pressure is more
important than galactic winds in the center. The interstellar material
in a galaxy feels the ram-pressure of the intracluster medium as it
flows past. This ram-pressure is
\begin{equation}\label{eq:rampressure}
P_r\propto \rho_{ICM} v^2,
\end{equation}   
where $\rho_{ICM}$ is the ICM density and $v$ is the relative velocity
between the galaxy and the ICM. \\ From the virial theorem we know
that the velocity of the galaxies is related to the total mass of the
cluster. A higher mass of cluster will translate to a higher velocity
of galaxy and due to the Eq. \ref{eq:rampressure} to a higher metal mass
as a consequence of ram-pressure stripping.
\cite{2004A&A...419....7D} found that the iron mass
associated with the abundance excess does not favor a scenario where
the iron mass is accreted from the cooling flow and that the excess
can be entirely produced by the brightest cluster galaxy (BCG) at the
centre of cool core clusters. On the other hand in our sample of 5
clusters we found that in the very central part, at an overdensity of
4500, the metal mass seems to be correlated with the total mass. The
clusters with higher mass are able to strip more gas from the galaxies
and to explain at least part of the central mass.

\section{Conclusions}
Based on \xmm observations, we studied the spatial distribution of
metal abundances in a sample of 5 relaxed clusters. Below we summarize
the main results.

\begin{itemize}
      \item Even for relaxed clusters the distribution of metals is
        clearly non-spherical. It looks very inhomogeneous with
        several maxima separated by low metallicity regions.
      \item The deviation from the expected temperature-metallicity
        relation suggests that several processes are at work in galaxy
        clusters and that the simple picture of stripped gas does not
        hold.
      \item The radial profiles is consistent with the current idea
        that the relative contribution of SN Ia to SNCC increases
        towards the cluster center and the O/Fe is more sensitive to
        this ratio than Si/Fe.
      \item Varying from cluster to cluster, the relative number of
        core-collapse supernovae necessary to reproduce the observed
        abundances ranges between 65-80$\%$
      \item Using a single extraction region to determine the
        metallicity gives a systematic underestimation of the metal
        mass: the metal masses are typically understimated by
        10-30$\%$.
\end{itemize}

\begin{acknowledgements}
      Authors are grateful to R. Paladino and C. Ferrari for useful
      comments and discussions, and the anonymous referee for his/her
      constructive suggestions.  We thank M. Murgia for the use of
      the SYNAGE++ program.  This work was supported by the Austrian
      Science Foundation (FWF) through grants P18523-N16 and
      P19300-N16.
\end{acknowledgements}

%\begin{thebibliography}{}

%\end{thebibliography}
\bibliographystyle{aa} \bibliography{paper}

%\Online

\end{document}